\documentclass[sigconf]{acmart}
\AtBeginDocument{%
  \providecommand\BibTeX{{%
    \normalfont B\kern-0.5em{\scshape i\kern-0.25em b}\kern-0.8em\TeX}}}

\setcopyright{acmlicensed}
\copyrightyear{2018}
\acmYear{2018}
\acmDOI{XXXXXXX.XXXXXXX}

\acmConference[Conference acronym 'XX]{Make sure to enter the correct
  conference title from your rights confirmation emai}{June 03--05,
  2018}{Woodstock, NY}
\acmBooktitle{Woodstock '18: ACM Symposium on Neural Gaze Detection,
 June 03--05, 2018, Woodstock, NY} 
\acmISBN{978-1-4503-XXXX-X/18/06}

\usepackage{multirow}
\usepackage{mdframed}
\usepackage{framed,enumitem} 
\usepackage{graphicx}
\usepackage{xspace}
\usepackage{makecell}
\usepackage{soul}
\usepackage{xcolor}
\usepackage[table]{xcolor}
\usepackage{textcomp}
\usepackage{array}
\usepackage{longtable}
\usepackage{tabularx}
\usepackage{listings}
\usepackage{subcaption}
\usepackage{booktabs}
\definecolor{mygreen}{RGB}{0,150,0}
\definecolor{myblue}{RGB}{30,100,200}
\definecolor{myred}{RGB}{180,0,0}
\definecolor{myyellow}{RGB}{200,150,0}
\definecolor{tabnote}{RGB}{178,34,34}
\setulcolor{mygreen}                
\setul{0.3ex}{0.3ex}   

\newcommand{\totalTurnN}{139}

\newcommand{\annotatedReasoningStepN}{992}
\newcommand{\reasoningStepN}{956}

\definecolor{blockcolor}{HTML}{000000}
\definecolor{blockrule}{gray}{0.6}
\newmdenv[
  topline=false,
  bottomline=false,
  rightline=false,
  leftline=true,
  linecolor=blockrule,
  linewidth=1pt,
  innertopmargin=2pt,
  innerbottommargin=2pt,
  innerleftmargin=10pt,
  innerrightmargin=10pt,
  skipabove=10pt,
  skipbelow=10pt,
  backgroundcolor=white,
  font=\small\color{blockcolor}
]{block}

\newmdenv[
  topline=true,
  bottomline=true,
  rightline=true,
  leftline=true,
  linecolor=blockrule,
  linewidth=1pt,
  innertopmargin=5pt,
  innerbottommargin=5pt,
  innerleftmargin=5pt,
  innerrightmargin=5pt,
  skipabove=5pt,
  skipbelow=10pt,
  backgroundcolor=white,
  font=\sffamily\color{blockcolor},
]{task}

\lstnewenvironment{prompt}{%
  \lstset{
    breaklines=true,
    linewidth=\linewidth,
    columns=fullflexible,
    keepspaces=true,
    basicstyle=\ttfamily\footnotesize,
    frame=single
  }
}{}

\begin{document}


% \title{What Did the AI Do on Its Own? Characterizing LLM Cognitive Work and User Delegation}
\title{What Did the AI Take On? Characterizing Cognitive Delegation in LLM Reasoning}

\author{Yoonsu Kim}
\orcid{0000-0002-9782-086X}
\affiliation{\institution{School of Computing \\ KAIST}
\city{Daejeon}
\country{Republic of Korea}}
\email{yoonsu16@kaist.ac.kr}

\author{Sean Kim}
\orcid{0009-0001-7707-1822}
\affiliation{\institution{School of Computing \\ KAIST}
\city{Daejeon}
\country{Republic of Korea}}
\email{kimsean@kaist.ac.kr}

\author{Kihoon Son}
\orcid{0000-0001-7224-2947}
\affiliation{\institution{School of Computing \\ KAIST}
\city{Daejeon}
\country{Republic of Korea}}
\email{kihoon.son@kaist.ac.kr}

\author{Saelyne Yang}
\orcid{0000-0003-1776-4712}
\affiliation{\institution{Carnegie Mellon University}
\city{Pittsburgh, PA}
\country{USA}}
\email{saelyney@andrew.cmu.edu}

\author{Juho Kim}
\email{juhokim@kaist.ac.kr}
\orcid{0000-0001-6348-4127}
\affiliation{
    \institution{School of Computing, KAIST}
    \city{Daejeon}
    \country{Republic of Korea}
}
\email{juho@skillbench.com}
\affiliation{
    \institution{SkillBench}
    \city{Santa Barbara, CA}
    \country{USA}
}

\renewcommand{\shortauthors}{Yoonsu Kim et al.}


% \ccsdesc[300]{Human-centered computing~HCI design and evaluation methods}



% %% A "teaser" image appears between the author and affiliation
% %% information and the body of the document, and typically spans the
% TEASER FIGURE
\begin{teaserfigure}
\end{teaserfigure}

\begin{abstract}
Large language models (LLMs) often perform intermediate cognitive work while carrying out users' requests, yet it remains unclear which parts users intended to delegate and how they wanted to remain involved. This matters because consequential choices may go unnoticed, limiting users' ability to steer the process, while reviewing every step would make delegation burdensome. We examined this with 24 LLM users across three knowledge-work tasks, collecting 992 retrospective annotations of reasoning steps. From this, we developed taxonomies of LLM cognitive work, delegation enactment, and desired delegation protocols at the reasoning-step level. Our analysis revealed that participants viewed about half of all steps (48.6\%) as AI-initiated, meaning the AI took on work they had not requested. Desired involvement varied with cognitive work and delegation enactment, even when contributions matched participants' intent. We propose design implications and sketches for supporting more deliberate cognitive delegation through flexible protocols and inspectable, revisable AI-initiated decisions.

% LLMs often perform intermediate cognitive work while carrying out users’ requests, yet it remains unclear which parts of this work users have intended to delegate to the AI. 
% Delegating a task to an AI also delegates the cognitive work required to carry it out. Yet not all of that work is explicitly handed over by the user. We investigate how this cognitive delegation unfolds within LLM reasoning: what cognitive work the model performs, who initiates each piece, and how users would have preferred the delegation to proceed. In our study, 24 participants completed one of three knowledge-work tasks and retrospectively annotated 992 AI's reasoning steps. We developed taxonomies of LLM cognitive work, delegation enactment, and desired delegation protocols. Our analysis revealed that 48.6\% of steps were AI-initiated, with AI initiative increasing from 31\% to 61\% within a turn and concentrating on unwanted and minor work. Desired protocols also varied across cognitive work and delegation enactment. From these findings, we propose design implications and sketches for supporting more deliberate cognitive delegation.

\end{abstract}
\maketitle
\section{Introduction}
People often delegate a task to an LLM through a prompt without specifying every piece of cognitive work required to carry it out~\cite{subramonyam2024gulf, nielsen2023ai}.
Reasoning-capable LLM systems can turn a task request into a sequence of intermediate reasoning steps and tool actions, forming and revising an approach as they work toward an output~\cite{yao2023react,Pang2026Interactive}. In doing so, they may resolve choices the user left unstated---what assumptions to make, which possibilities to pursue, and how much detail to develop---by inferring defaults for unspecified requirements~\cite{shaikh-etal-2024-grounding,yang2026prompts}. Such an AI initiative is part of what makes delegation useful. However, asking an LLM to complete a task does not necessarily specify how the user wants every intermediate decision to be handled. This raises the question of which decisions users want to inspect or revise, and when they want to be involved.

Prior systems address this by giving users oversight of AI outputs and how they are produced, for example by exposing plans, reasoning steps, or contribution provenance~\cite{ghosh2026experimental,wu2022aichains,Pang2026Interactive,hoque2024hallmark}. Yet such oversight does not by itself indicate which parts of AI-initiated work require users' attention. Inspecting or confirming every step would make delegation burdensome~\cite{Pang2026Interactive,bucinca2021trust}. Studies of delegation further show that desired involvement varies across tasks and contexts, but have examined these preferences at the level of tasks, workflows, or interaction modes~\cite{song2026good,khurana2025do,he2025plan}. What remains unclear is a user-grounded understanding of how delegation unfolds across the intermediate work of producing a single response: what users intended the AI to take on and how they wanted to be involved.

We studied this at the level of reasoning steps and tool actions in a study with 24 frequent LLM users. Each participant completed one of three predefined tasks, each representing a major category of LLM-assisted knowledge work identified in prior work: Creation, Information, or Advice~\cite{Lee2025Impact,Brachman2024Knowledge}. Using a custom interface for task completion and retrospective review, we collected \annotatedReasoningStepN{} step-level annotations in which participants judged, looking back at each step, whether they had asked for it or the AI had taken it on by itself, how critical it was, whether its contribution matched their intent, and how they would have wanted to be involved. From these data, we developed three taxonomies at the level of LLM reasoning steps, characterizing the cognitive work each step performed, how participants perceived that work to have been delegated, and how they wanted the delegation to proceed. Because all three apply to the same steps, we can examine how the kind of work relates to participants' interpretation of its delegation and their preferred coordination.

Our step-level analysis revealed patterns obscured at the task level. 
Participants judged 48.6\% of reasoning steps as AI-initiated, meaning they saw the AI as taking on work they had not asked it to perform. This share rose from 31.1\% in the first fifth of a turn to 61.4\% in the final fifth. Yet AI initiative was not necessarily unwanted: 62.7\% of AI-initiated steps contributed as participants wanted. Even so, unwanted and minor steps were more often AI-initiated than wanted and critical steps (70.0\% vs.\ 41.1\%; 75.8\% vs.\ 44.4\%).
Participants did not want oversight of everything, preferring full delegation for 63.9\% of all steps. However, they wanted consultation when the AI took on critical work they had not intended to delegate, and less AI involvement when it performed unnecessary minor work. They also wanted involvement in 28\% of critical AI-initiated steps that contributed as they wanted, including most requests for after-the-fact disclosure. Desired involvement also varied by cognitive work: consultation for \textit{Generating}, calibration toward doing less for \textit{Repeating}, and relatively little involvement for \textit{Understanding}, \textit{Planning}, and \textit{Recalling}. Together, these findings suggest supporting selective involvement in individual decisions throughout a response, including when AI initiative yields contributions that participants value.

Grounded in these findings, we propose two design implications for deliberate cognitive delegation and explore them through low-fidelity sketches. First, interfaces should support multiple delegation protocols within a response to accommodate different involvement preferences across cognitive work. Second, they should make consequential AI-initiated decisions inspectable and revisable, addressing the desire for involvement even when contributions are satisfactory.

We make the following contributions:
\begin{itemize}
    \item Three complementary, user-grounded taxonomies that characterize the cognitive work performed in LLM reasoning, how its delegation is enacted, and how users want that delegation to proceed.
    \item An empirical analysis of \annotatedReasoningStepN{} annotated step-level instances from 24 participants across three knowledge-work tasks, revealing patterns in perceived AI initiative, alignment with user intentions, and desired involvement.
    \item Two design implications, explored through low-fidelity sketches, for tailoring delegation protocols and inspecting and revising AI-initiated decisions.
\end{itemize}

\section{Related Work}
We position our work at the level of the intermediate cognitive work an LLM performs within a single response, extending prior research on AI delegation, reliance, and the division of work in human--AI collaboration.
% We position our work at the level of the individual cognitive steps through which an LLM carries out delegated work, moving beyond prior work that primarily focuses on whether users delegate an entire task, follow an AI recommendation, or interact with the system.

\subsection{Delegation, Reliance, and User Oversight in Human--AI Collaboration}
AI systems can support a wide range of knowledge work, including information seeking, analysis, and artifact creation~\cite{Brachman2024Knowledge}. Delegating this work can reduce users' workload~\cite{padmakumar2026offloading}, but extensive or unreflective reliance may weaken their oversight of how a task is carried out. Prior work also finds that extensive AI assistance can reduce users' persistence and cognitive effort and hinder skill formation~\cite{liu2026assistance,Lee2025Impact,shen2026skill,caosun2026augmentation}. When users delegate tasks end-to-end, they may fail to notice when outcomes degrade as work unfolds~\cite{laban2026corrupt}. At the extreme, this can become ``cognitive surrender,'' in which users relinquish responsibility for consequential cognitive work~\cite{shaw2026thinking}.

Research on AI delegation has examined how users decide whether and what to delegate to AI~\cite{spitzer2025human}. Related work on appropriate reliance examines whether users appropriately accept or reject AI recommendations~\cite{passi2022overreliance}. Prior systems have explored how interfaces can help users assess these recommendations and calibrate their reliance~\cite{passi2024appropriate}. More broadly, this literature distinguishes reliance from decision quality, where following or rejecting AI advice can each be appropriate or inappropriate depending on its correctness~\cite{schoeffer2025ai}.

However, it remains unclear how delegation unfolds across the intermediate work an LLM performs while generating a single response to a user's request. A task request does not show which work users intended the AI to take on or how they wanted to be involved. We address this gap by linking step-level cognitive work with users' interpretations of its delegation and their preferred involvement.

\subsection{Division of Cognitive Work in Human--AI Collaboration}
Characterizing delegation within a single LLM response requires distinguishing what cognitive work is performed from how that work is divided between users and AI. 
Function-allocation research provides a basis for this distinction by separating the type of function being automated, such as information acquisition, analysis, decision selection, or action, from its level of automation~\cite{parasuraman2000model}.
% Work on mixed-initiative interfaces similarly proposes that systems should decide whether to act, ask, or defer by considering uncertainty about user goals and the cost of incorrect autonomous action~\cite{horvitz1999principles}.
Research on adaptive autonomy further shows that no single, fixed level of AI control is optimal across situations~\cite{salikutluk2024evaluation}.
Together, this work frames delegation as a coordination problem involving different functions and shifting degrees of initiative, rather than a binary choice between manual work and full automation.

Research on generative AI has extended these ideas to knowledge-work domains. 
Studies of LLM-supported learning, writing, design ideation, and professional work show that users delegate different parts of a task and that control can shift between users and AI as work progresses~\cite{song2026good, wan2024felt, nguyen2024human, shi2026taxonomy, ma2024human, Brachman2024Knowledge}.
Broader taxonomies likewise describe human-AI interaction in terms of stages such as planning, facilitating, iterating, and testing~\cite{gao2024taxonomy}, or dimensions such as agency, interaction, and adaptation~\cite{holter2024deconstructing}.
However, these accounts generally operate at the level of tasks, stages, or interaction modes. Finer-grained studies examine how programmers respond to code suggestions or shift between acceleration and exploration, but still focus on work visible in the interaction between the user and the system~\cite{mozannar2024reading,barke2023grounded}.

Contemporary reasoning models can carry out multiple reasoning steps and tool actions in response to a single prompt, making consequential decisions without a corresponding user-visible exchange.
% As a result, an interaction that appears user-led at the conversational level may still contain substantial AI-led cognitive work within the task.
% Existing work therefore does not explain how this hidden, within-task cognitive work is actually divided between users and AI.
Existing task- and interaction-level accounts therefore offer limited insight into how cognitive work is divided within a single response. We address this gap by relating the cognitive work performed at each reasoning step to how users perceived its delegation.

\subsection{Supporting User Involvement of AI-Initiated Cognitive Work}
When an LLM can make consequential decisions within a task without a user-visible exchange, an important design question is how users can become aware of and retain control over this AI-initiated cognitive work.
Prior research has proposed several approaches to help users evaluate, inspect, and intervene in AI assistance.

One line of work supports more deliberate evaluation of AI recommendations through calibrated uncertainty~\cite{cao2024designing}, sources and inconsistencies in LLM explanations~\cite{kim2025fostering}, or cognitive forcing functions~\cite{bucinca2021trust}. These mechanisms help users decide whether to accept an output but offer limited visibility into the intermediate work that produced it.

A second line of work gives users visibility into and influence over how AI carries out a task. Before execution, AI-generated plans allow users to review and revise the proposed course of action~\cite{ghosh2026experimental}, while interactive controls surface underspecified choices in a request~\cite{zhang2026words}. Guided copilots instead give users opportunities to participate as the task progresses, rather than delegating it end-to-end~\cite{khurana2025do}.
Other systems support inspection during or after execution. Editable workflows expose intermediate steps and results that users can modify~\cite{wu2022aichains}, while hierarchical representations help users navigate and revise model-produced reasoning~\cite{Pang2026Interactive}. Provenance views show how human and AI contributions combine to shape a final artifact~\cite{hoque2024hallmark}.

These systems provide different points of intervention but offer limited guidance on how appropriate involvement depends on the cognitive work performed or how users perceive its delegation. We address this by relating users' desired involvement to both dimensions at the reasoning-step level, informing design implications for deliberate cognitive delegation.

\section{Study Design}
We designed a user study to characterize how cognitive work is distributed between users and LLMs during knowledge-work tasks. In particular, we focused on the individual reasoning steps through which an LLM carries out a task that was delegated by the user, how delegation unfolds within those steps, and how users would prefer that delegation to proceed.

\subsection{Tasks}
We studied LLM-assisted knowledge work across three major usage categories identified in prior HCI research: \textbf{Creation},
\textbf{Information}, and \textbf{Advice}~\cite{Lee2025Impact, Brachman2024Knowledge} (\autoref{tab:llm-task-categories}).
\begin{table}[t]
    \centering

    \begin{tabularx}{\linewidth}{@{}
        >{\raggedright\arraybackslash}p{0.13\linewidth}
        >{\raggedright\arraybackslash}p{0.22\linewidth}
        >{\raggedright\arraybackslash}X
    @{}}
        \toprule
        \textbf{Category} & \textbf{Sub-category} & \textbf{Description} \\
        \midrule
        \textbf{Creation}
        & Idea
        & Generate an idea, to be used indirectly \\
        & Artifact
        & Generate a new artifact to be used directly or with some modification \\
        \midrule
        \textbf{Infor\-mation}
        & Search
        & Seek a fact or piece of information \\
        & Learn
        & Learn about a new topic more broadly \\
        & Summarize
        & Generate a shorter version of a piece of content that describes the important elements \\
        & Analyze
        & Discover a new insight about information or data \\
        \midrule
        \textbf{Advice}
        & Validation
        & Check whether an artifact satisfies a set of rules or constraints \\
        & Guidance
        & Get guidance about how to make a decision \\
        & Improve
        & Generate a better version \\
        \bottomrule
    \end{tabularx}
    \caption{Knowledge-work categories and sub-categories covered in our study.}
    \label{tab:llm-task-categories}
    \Description{A three-column table listing knowledge-work categories, sub-categories, and descriptions used in the study. The categories are Creation, Information, and Advice. Creation includes Artifact and Idea; Information includes Search, Learn, Summarize, and Analyze; Advice includes Improve, Guidance, and Validation. Each row briefly defines the purpose of the corresponding sub-category.}
\end{table}

We included all three categories to examine how cognitive work and delegation unfold in practice across different types of LLM-assisted knowledge work.
We designed one task for each category. Each task was intentionally designed to cover all of the sub-categories within its corresponding category, rather than focusing on only one sub-category.

\begin{itemize}
    \item \textbf{Creation: creating pitch slides through ideation.} Participants acted as the founder of an early-stage startup preparing a pitch to early investors. We gave them a theme, micro-wellbeing services for busy people, and left the rest open: which everyday moments to address, what form the service should take, and whom to target. They first \textbf{\textit{generated an idea}} for a service and explored possible directions before settling on one. They then \textbf{\textit{generated an artifact}} by turning the selected concept into a pitch deck, with no constraint on its length or structure. 
    \item \textbf{Information: writing an explainer.} Participants acted as a writer preparing an explainer for a general audience comparing three modes of electricity generation: renewables, nuclear, and fossil fuels. The intended reader was assumed to know these terms but not how they work. Participants \textbf{\textit{searched}} for facts and data on the three modes, \textbf{\textit{learned}} how each works and what characteristics it has, \textbf{\textit{summarized}} their findings down to what mattered, and \textbf{\textit{analyzed}} the material to surface points that emerge only from comparison. 
    \item \textbf{Advice: revising a cover letter.} Participants acted as a senior colleague asked by a junior one to revise a cover letter for a marketing position at a food company. We provided the draft, which contained planted weaknesses, along with the job posting. Participants \textbf{\emph{validated}} the draft against the posting's requirements, formed \textbf{\emph{guidance}} on which problems mattered and in what direction to fix them, and \textbf{\emph{improved}} the letter accordingly. These activities recurred as participants assessed and revised the letter.
\end{itemize}
\begin{figure*}
    \centering
    \includegraphics[width=1\linewidth]{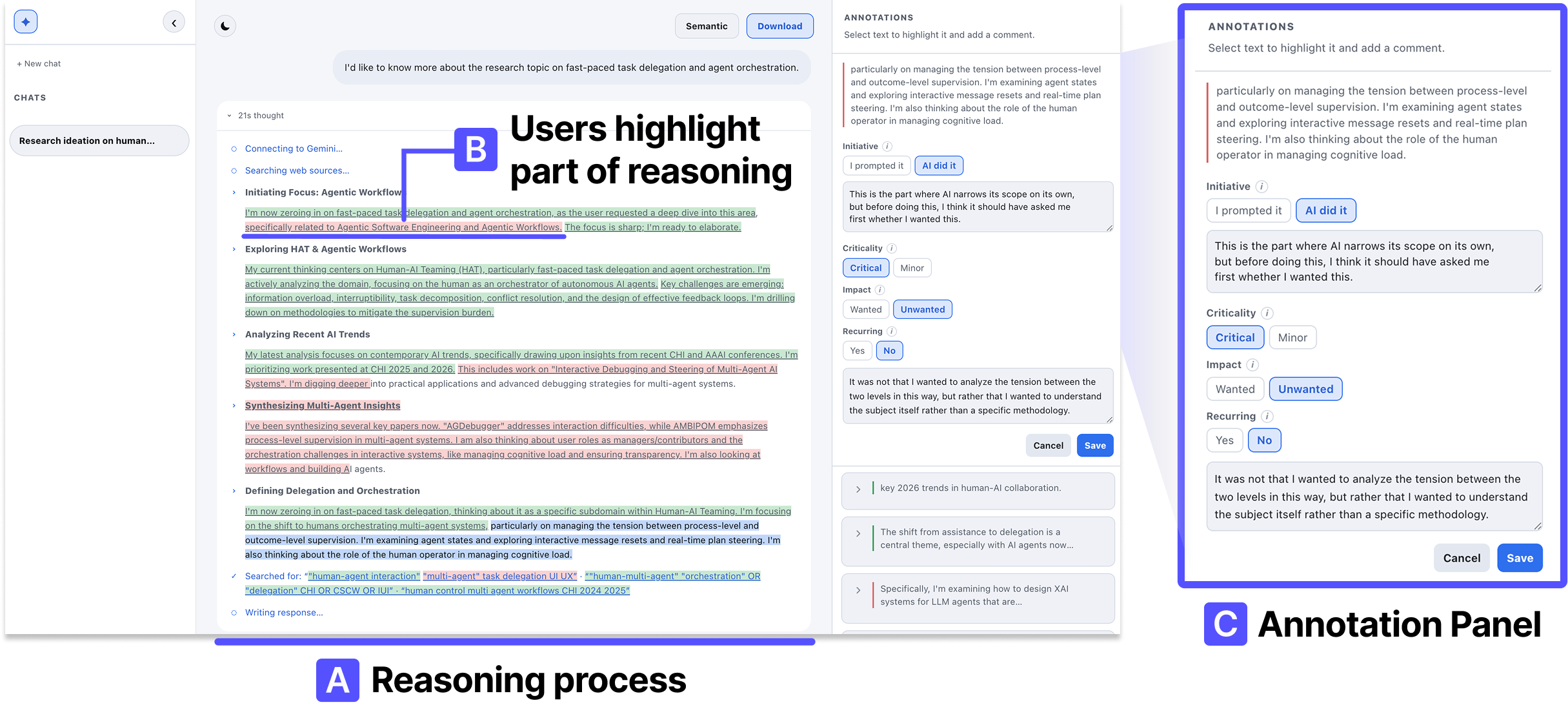}
    \caption{Interface for inspecting and annotating an LLM’s reasoning trace. (A) The main workspace shows the model’s multi-step reasoning as it responds to a user’s prompt. (B) The user selects a specific span of the reasoning they want to annotate. (C) The annotation panel lets users label that span along multiple dimensions (initiative, criticality, impact, recurrence) and add free-text comments about how it relates to what they meant to delegate.}
    \Description{Screenshot of a custom interface for reviewing and annotating an AI assistant’s reasoning. On the left, a chat-like panel displays a model-produced multi-step reasoning trace for a user prompt. On the right, an annotation panel lets the user label the selected reasoning step text as AI-initiated or user-initiated, as Critical or Minor, and as Wanted or Unwanted based on its contribution to the reviewed output, mark whether this kind of step is recurring, and add a free-text comment.}
    \label{fig:interface}
\end{figure*}

\subsection{LLM Reasoning-Trace Inspection Interface (\autoref{fig:interface})}
We developed a custom interface for two purposes: LLM-assisted task completion and retrospective inspection and annotation of the reasoning steps and tool actions underlying each response. The interface integrated a reasoning-capable LLM (\texttt{gemini-3.5-flash}) with web search and a PowerPoint-generation tool implemented through the Model Context Protocol (MCP).
For each turn, the system logged the user's prompt, the model's reasoning steps and tool actions, and its final output. After completing the task, participants revisited this interaction history through the same interface. They could inspect the reasoning steps and tool actions underlying each AI output, select specific spans, and annotate them using structured labels and free-form comments.
To reduce the burden of entering detailed qualitative feedback, the interface also supported speech-to-text input. Participants could verbalize their explanations while annotating a reasoning step, and the system transcribed them into the free-text annotation field. Participants could review and edit each transcription before submitting it. 

\subsection{Participants}
We recruited 24 participants (12 female, 12 male; age $M = 25.75$, $SD = 2.85$) through recruitment postings on our university's online community platform. Participants were frequent users of LLM-based AI assistants such as ChatGPT, Claude, and Gemini ($M = 4.42$, $SD = 0.83$ on a 5-point frequency scale, 1 = no prior use, 5 = multiple times a day). We assigned participants to one of the three tasks based on their prior experience with each task, ensuring that no participant worked on a task that was unfamiliar to them ($M = 4.00$, $SD = 0.74$ on a 5-point scale, 1 = no experience, 5 = very frequent experience). Eight participants were assigned to each task, and each participant received KRW 40,000 ($\approx$ USD 29) for participating in the approximately two-hour study.

\subsection{Procedure}
Each study session lasted approximately two hours and consisted of three stages: (1) a 10-minute introduction covering informed consent, task instructions, and the LLM interface; (2) a 20-minute \textbf{task phase}, in which participants completed their assigned task with the LLM; and 
(3) a 90-minute \textbf{retrospective annotation and interview phase}, in which participants reviewed the recorded trace of how the LLM carried out their prompts, including its reasoning steps and tool-use actions, annotated selected parts of the trace, and participated in a semi-structured interview.
The study was approved by our institution's IRB, and all participants provided informed consent before participating in the study.

\subsubsection{Task Phase}
After the introduction to their assigned task and the LLM interface, participants were instructed to complete the task using the study system as they normally would. We provided no guidance on how to formulate prompts, which parts of the task to ask the AI to handle, or how to structure their interaction with it. 
Participants were therefore free to decide what to ask the AI to do, what to retain for themselves, and how much autonomy to allow the AI during the task. The complete interaction history, including user prompts, AI reasoning traces, tool calls, and outputs, was recorded for the subsequent retrospective annotation phase.

\subsubsection{Retrospective Reasoning Annotation and Interview Phase}
After completing the task, participants revisited their recorded interaction turn by turn. Before inspecting the AI's reasoning and tool-use traces, they described how they understood their own and the AI's roles in completing the task.
For each turn, participants first recalled their intent and reviewed the corresponding AI output, marking notable good or problematic parts. This output-level review provided context for subsequently inspecting the associated AI reasoning and tool-use trace, but was not itself a primary unit of analysis.
Participants then inspected each reasoning step and tool-use action, providing at least one annotation per step or action. They could add further annotations to notable spans and were encouraged to connect parts of the output to the work that may have produced them. For each annotation, participants made three structured judgments: \emph{initiative}, \emph{criticality}, and \emph{impact}. We selected these dimensions to distinguish whether each step reflected work the user had intended to delegate or work the AI took on without being asked, how consequential that work was, and whether its contribution ultimately aligned with the user's desired outcome. Treating these as separate judgments allowed us to distinguish how delegation was enacted from whether it was successful~\cite{he2023rebalancing,subramonyam2024gulf}.

\begin{itemize}
    \item \textbf{Initiative: who set this work in motion?}
    Participants judged whether the work was something they had prompted the AI to do as (``I prompted it'' -- \emph{User-initiated}) or something the AI had done on its own (``AI did it'' -- \emph{AI-initiated}).
    \item \textbf{Criticality: how consequential was this work?}
    Participants judged whether the work was \emph{Critical} or \emph{Minor} to the overall task and its outcome.
    \item \textbf{Impact: did this work contribute to an outcome aligned with what users wanted?}
    Participants judged whether its contribution was \emph{Wanted} or \emph{Unwanted}, relating the step or action to the output they had reviewed.
\end{itemize}
For every annotation, participants explained each judgment in their own words. These free-form comments provided qualitative evidence for understanding why participants made each judgment. Participants typically verbalized their explanations while annotating; the speech-recognition system transcribed them into text, which participants could review and edit before submission. In a subsequent semi-structured interview, participants discussed how they perceived work the AI had taken on without being asked, how they wished to be involved in such work, and how they would prefer to recover when it produced unwanted outcomes.

\subsection{Data Collection and Dataset Construction}
Our dataset combines recorded LLM interactions with participants' retrospective annotations and qualitative feedback. We collected all user prompts, AI outputs, reasoning steps, and tool-use actions, along with structured annotations, free-form comments, and screen and audio recordings.
% We used annotated reasoning steps as the primary unit of analysis, each representing a distinct piece of work performed by the AI. Participants annotated every step at least once and could add separate annotations for portions warranting different judgments. Tool-use actions were annotated in the same manner and are collectively referred to as reasoning steps.

We used step-level annotations as the primary unit of analysis. Each annotation captured participants' judgments about a distinct piece of work within an AI reasoning step or tool-use action. Across 24 participants, the study produced \reasoningStepN{} reasoning steps across \totalTurnN{} interaction turns ($M = 5.8$ turns per participant), yielding \annotatedReasoningStepN{} step-level annotations with free-form explanations ($M = 41.3$ annotations per participant). These annotations formed the basis for the taxonomies and analyses reported in the following sections.

\section{Taxonomies of LLM Cognitive Work and Delegation}\label{sec:taxonomy}
Through iterative analysis of participants' annotations and comments, we developed three taxonomies that describe each reasoning step or tool action along three dimensions: (1) what cognitive work the LLM performed, (2)  how delegation unfolded---whether the user had asked for the work or the AI took it on, and whether the contribution was what the user wanted, and (3) how the user wanted the AI to proceed. 
Together, the taxonomies let us examine what work the AI performed within a response, whether users had asked for it, and when and how they wanted to be involved.
% To characterize what occurs within an LLM's reasoning process when carrying out a user-delegated task, we developed a set of complementary taxonomies from the annotated reasoning traces. Our analysis distinguishes three aspects of each reasoning step. First, we characterize \emph{what kind of cognitive work} the LLM was performing. Second, we characterize \emph{how delegation was enacted} at that step---whether the work was explicitly prompted by the user or taken on by the AI, and whether this matched what the user wanted. Finally, we characterize \emph{how users wanted the delegation to proceed}, based on their free-form feedback about when and how they wanted to be involved.

% Together, these taxonomies provide complementary lenses on the same unit of analysis: an individual reasoning step and tool call action. This allows us to subsequently examine not only what work LLMs perform, but which forms of work they take on autonomously, when such autonomy aligns with or conflicts with users' intentions, and which forms of delegation users prefer for different kinds of work.

\subsection{Taxonomy Development}
We developed the taxonomies through an iterative qualitative analysis of the annotated reasoning steps and participants' accompanying free-form explanations.
The structured annotations captured participants' retrospective judgments of each step's initiative, criticality, and impact, while their free-form comments clarified these judgments and how they thought the AI should have proceeded.
Two authors first independently open-coded~\cite {gibbs2018analyzing} 247 data points (24.9\% of the full dataset) from six participants, balanced across the three tasks. For each instance, we developed initial codes around three questions: (1) what kind of cognitive work the AI was performing, (2) how that work came to be delegated to the AI and whether this aligned with the participant's intent, and (3) how the participant wanted the delegation to proceed.
We then brought the independently developed codes together through iterative affinity diagramming, grouping similar concepts, merging overlapping codes, and refining their definitions and boundaries through discussion. Through this process, we developed an initial set of codes that captured distinct patterns observed across the data. We next applied these initial codes to an additional 320 data points from six other participants, again with two participants from each task. The two authors independently coded these data and refined the code set when new cases were not adequately captured. This iterative process continued until the codes were stabilized across the 567 development data points.
To provide an independent review of the resulting structure, two additional authors joined the analysis to review the resulting codes and conduct axial coding~\cite{gibbs2018analyzing}, organizing related codes into higher-level categories and clarifying their relationships.
This process resulted in three taxonomies: (1) \textbf{LLM's Cognitive Work}, capturing the types of cognitive work performed within LLM reasoning steps and tool-use actions; (2) \textbf{Delegation Enactment}, capturing how that work came to be delegated to the AI and whether it aligned with the user's intent; and (3) \textbf{Desired Delegation Protocol}, capturing how users wanted the AI to communicate and proceed when performing that work.
Note that the \textbf{Delegation Enactment} taxonomy was derived by synthesizing participants' structured judgments of initiative, criticality, and alignment with user intent together with their accompanying explanations. We organized recurring combinations of these judgments into distinct patterns of delegation enactment.

\subsection{Taxonomy of Cognitive Work in LLM Reasoning Steps}
\begin{table*}[t]
\centering
\footnotesize
\begin{tabularx}{\textwidth}{
    >{\raggedright\arraybackslash}p{0.11\textwidth}
    >{\raggedright\arraybackslash}p{0.23\textwidth}
    >{\raggedright\arraybackslash}p{0.12\textwidth}
    >{\raggedright\arraybackslash}X
}
\toprule
\textbf{Category} &
\textbf{Category Description} &
\textbf{Code} &
\textbf{Code Description} \\
\midrule
\multirow{2}{*}{\textbf{Framing}} &
\multirow{2}{0.23\textwidth}{Establishing what to address and how to approach the task based on the user's prompt.} &
\textbf{Understanding} &
Interpreting the user's intent, task goal, and relevant context from the prompt. \\
\cmidrule(lr){3-4} &
&
\textbf{Planning} &
Formulating a strategy or sequence of actions for carrying out the task based on the user's prompt. \\
\midrule
\multirow{2}{*}{\textbf{Sourcing}} &
\multirow{2}{0.23\textwidth}{Bringing information needed for the task into the current reasoning process from existing sources.} &
\textbf{Gathering} &
Obtaining task-relevant information from external sources, such as through web search. \\
\cmidrule(lr){3-4} &
&
\textbf{Recalling} &
Retrieving relevant information from the preceding conversational context and bringing it into the current turn. \\
\midrule
\multirow{4}{*}{\textbf{Developing}} &
\multirow{4}{0.23\textwidth}{Directly developing the substantive content of the task by expanding, examining, narrowing, or creating it.} &
\textbf{Brainstorming} &
Exploring multiple possible ideas or directions generatively, rather than examining a single existing option. \\
\cmidrule(lr){3-4} &
&
\textbf{Analyzing} &
Examining information in depth, decomposing it into meaningful elements, comparing them, or deriving relationships from existing evidence. \\
\cmidrule(lr){3-4} &
&
\textbf{Narrowing} &
Selecting, prioritizing, or emphasizing particular directions or elements among multiple possibilities. \\
\cmidrule(lr){3-4} &
&
\textbf{Generating} &
Creating new substantive content or task artifacts, including artifact-generating tool actions, rather than merely restructuring existing content. \\
\midrule
\multirow{2}{*}{\textbf{Reviewing}} &
\multirow{2}{0.23\textwidth}{Meta-level work that examines and consolidates the AI's preceding reasoning.} &
\textbf{Evaluating} &
Assessing preceding reasoning or intermediate work against relevant criteria. \\
\cmidrule(lr){3-4} &
&
\textbf{Synthesizing} &
Integrating and consolidating multiple preceding reasoning steps into a coherent summary, understanding, or conclusion. \\
\midrule[\heavyrulewidth]
% \multicolumn{4}{@{}>{\raggedright\arraybackslash}p{\dimexpr\textwidth-2\tabcolsep\relax}@{}}{%
% {\textit{$\dagger$ Note.} The four categories above classify a step by the object of its work. \emph{Reprocessing} is set apart on a different basis: it identifies steps that return to one specific earlier step rather than taking on new work, and takes precedence when both apply.}} \\ \addlinespace[3pt] \hline
\addlinespace[4pt] 
\multirow{2}{*}{\textbf{Reprocessing} $\dagger$} &
\multirow{2}{0.23\textwidth}{Returning to and further processing content from a specific earlier reasoning step, rather than taking on new work.} &
\textbf{Elaborating} &
Further developing content from a specific earlier reasoning step by adding detail, specificity, or refinement. \\
\cmidrule(lr){3-4} &
&
\textbf{Repeating} &
Restating or reproducing content from a specific earlier reasoning step with little or no substantive new development. \\
\bottomrule
\end{tabularx}
% \vspace{3pt}
% \noindent\parbox{\textwidth}{\footnotesize\textcolor{black}{%
% \textit{$\dagger$ Note.} \emph{Reprocessing} is defined on a different basis from the four categories above it: rather than the object of a step's work, it identifies a step's return to a specific earlier step. It takes precedence when both apply, with the earlier step's own category recorded alongside it.}}
\caption{Taxonomy of cognitive work in LLM reasoning steps, comprising five categories and twelve codes. $\dagger$ Note: \emph{Reprocessing} is defined on a different basis from the four categories above it: rather than the object of a step's work, it identifies a step's return to a specific earlier step.}
\Description{A four-column table listing categories, category descriptions, codes, and code descriptions for a taxonomy of cognitive work in LLM reasoning steps. Five categories are listed. Framing, establishing what to address and how to approach the task based on the user's prompt, contains Understanding and Planning. Sourcing, bringing information needed for the task into the current reasoning process from existing sources, contains Gathering and Recalling. Developing, directly developing the substantive content of the task, contains Brainstorming, Analyzing, Narrowing, and Generating. Reviewing, meta-level work that examines and consolidates the AI's preceding reasoning, contains Evaluating and Synthesizing. A heavier horizontal rule separates these four from the fifth category, Reprocessing, marked with a dagger, which covers returning to and further processing content from a specific earlier reasoning step and contains Elaborating and Repeating. The dagger is explained in the caption.}
\label{tab:taxonomy_cw}
\end{table*}
Our first taxonomy characterizes \emph{what kind of cognitive work} the LLM performs within each reasoning step (\autoref{tab:taxonomy_cw}).
Like prior functional taxonomies of reasoning steps, but extending beyond their focus on benchmark problem solving~\cite{belkhiter2026TRACES,bogdan2025thought}, ours captures user-facing knowledge work across Creation, Information, and Advice. We identified five higher-level categories: \emph{Framing}, \emph{Sourcing}, \emph{Developing}, \emph{Reviewing}, and \emph{Reprocessing}.
The first four characterize the primary object and function of a reasoning step. \emph{Reprocessing}, in contrast, captures cases in which the model revisits and further processes work from an earlier reasoning step.
We set it apart because participants perceived these steps differently. When looking at what cognitive work a reasoning step performed, they interpreted a step that returns to earlier work in relation to that earlier step rather than on its own. We coded the five categories as mutually exclusive. For \emph{Reprocessing} steps, we additionally recorded which earlier step they returned to and which category that step had received.

\smallskip
\noindent
\textbf{\textit{Framing.}}
It captures work that establishes how the task will be understood and approached based on the user's prompt. It includes \emph{Understanding}, in which the model interprets the user's intent, task goal, and relevant context, and \emph{Planning}, in which it formulates a strategy or sequence of actions for carrying out the task.

\smallskip
\noindent
\textbf{\textit{Sourcing.}}
It encompasses work that brings information needed for the task into the current reasoning process. \emph{Gathering} obtains information from external sources, such as through web search or tool use, whereas \emph{Recalling} retrieves relevant information from the preceding conversational context.

\smallskip
\noindent
\textbf{\textit{Developing.}}
It covers work that directly advances the substantive content of the task. It includes \emph{Brainstorming} alternative possibilities, \emph{Analyzing} information or decomposing it into meaningful elements, \emph{Narrowing} among alternatives or emphasizing selected dimensions, and \emph{Generating} new task artifacts or substantive content.

\smallskip
\noindent
\textbf{\textit{Reviewing.}}
It captures meta-level work in which the model examines its own preceding reasoning. \emph{Evaluating} assesses prior reasoning steps against relevant criteria, while \emph{Synthesizing} integrates multiple prior reasoning steps into a consolidated understanding or conclusion.

\smallskip
\noindent
\textbf{\textit{Reprocessing.}}
It characterizes cases in which the model returns to content from a specific earlier reasoning step. \emph{Elaborating} develops that earlier work with additional detail or specificity, whereas \emph{Repeating} restates or reproduces it without substantial new development.

\subsection{Taxonomy of Delegation Enactment in LLM Reasoning Steps}
\begin{table*}[t]
\centering
\small
\renewcommand{\arraystretch}{1.2}
\setlength{\tabcolsep}{4pt}
\setlength{\aboverulesep}{0pt}
\setlength{\belowrulesep}{0pt}
\setlength{\extrarowheight}{2pt}
\begin{tabular}{@{}p{1.7cm}|p{0.8cm}|p{5.6cm}|p{0.8cm}|p{5.6cm}@{}}
\toprule
& \multicolumn{2}{c|}{\textbf{Impact: Wanted}}
& \multicolumn{2}{c}{\textbf{Impact: Unwanted}} \\ 

\toprule

\textbf{User-Initiated}
& \multicolumn{2}{p{6.8cm}|}{\textbf{Intended Delegation} \newline
  The user explicitly prompted the AI to perform the work, and the AI's handling was aligned with what the user wanted.}
& \multicolumn{2}{p{6.8cm}}{\textbf{Regretted Delegation} \newline
  The user explicitly prompted the AI to perform the work, but its handling was not what the user wanted in retrospect.} \\ 

\toprule

\textbf{AI-Initiated}
& \multicolumn{2}{p{6.8cm}|}{\textbf{Serendipitous Delegation} \newline
  The user did not explicitly prompt or anticipate this specific work, but the AI took it on autonomously, and the user found its contribution to the output wanted or unexpectedly valuable.}
& \multicolumn{2}{p{6.8cm}}{\textbf{Unintended Delegation} \newline
  The user did not explicitly prompt this specific work, and the AI took it on autonomously in a way that was not what the user wanted.} \\ \cmidrule{2-5}

& \emph{Critical}
& \textbf{Beneficial Autonomy} \newline
  % The AI autonomously took on consequential work that the user found aligned with their intent and meaningfully valuable.
  The AI autonomously took on work that shaped the outcome in a way the user valued, contributing something beyond what the prompt had asked for. 
& \emph{Critical}
& \textbf{Overreaching Delegation} \newline
  The AI autonomously took on consequential work that the user wanted to retain control
  over or be involved in. \\ \cmidrule{2-5}

& \emph{Minor}
& \textbf{Routine Autonomy} \newline
  The AI autonomously handled minor or routine work that the user found appropriate and helpful.
& \emph{Minor}
& \textbf{Unnecessary Delegation} \newline
  The AI autonomously performed minor work that the user did not want or consider necessary. \\

\bottomrule
\end{tabular}
\caption{Taxonomy of Delegation Enactment. Crossing initiative (\emph{User-Initiated} / \emph{AI-Initiated}) with impact (\emph{Wanted} / \emph{Unwanted}) yields four forms of delegation enactment. Each AI-initiated form is further split by whether participants judged the work critical or minor to the overall task.}
\Description{A two-by-two table crossing initiative with impact. Rows are User-Initiated and AI-Initiated; columns are Impact: Wanted and Impact: Unwanted. The four cells are Intended Delegation, Regretted Delegation, Serendipitous Delegation, and Unintended Delegation, each with a one-sentence definition. The two AI-Initiated cells are each subdivided by criticality: Serendipitous Delegation into Beneficial Autonomy for critical work and Routine Autonomy for minor work, and Unintended Delegation into Overreaching Delegation for critical work and Unnecessary Delegation for minor work.}
\label{tab:delegation_enactment}
\end{table*}

Delegation begins with the user's prompt: by asking the LLM to complete a task, the user delegates that task at a high level. However, the user's prompt does not specify every piece of work required to carry it out~\cite{lam2026jit}, and the model can introduce additional operations, judgments, or decisions as it reasons toward the requested outcome. We therefore introduce \emph{delegation enactment} to describe how a high-level delegation is concretely carried out at the level of individual reasoning steps.
We characterize delegation enactment along two user-annotated dimensions: \emph{initiative} and \emph{impact}. Initiative distinguishes work that the user explicitly prompted, which we refer to as \emph{user-initiated}, from work that the AI independently took on while carrying out the user's prompt, which we refer to as \emph{AI-initiated}. Impact concerns what a step contributed to the resulting output: participants judged whether that contribution was \emph{wanted} or \emph{unwanted}. Crossing these dimensions yields four different forms of delegation enactment (\autoref{tab:delegation_enactment}). 
% \emph{Intended Delegation}, \emph{Regretted Delegation}, \emph{Serendipitous Delegation}, and \emph{Unintended Delegation} (\autoref{tab:delegation_enactment}). 

\smallskip
\noindent
\textbf{\textit{Intended Delegation.}}
It refers to work that the user explicitly asked the AI to perform and that the AI handled in accordance with the user's intent. It represents cases in which both the allocation of the work and its execution matched what the user intended to delegate.

\smallskip\noindent
\textbf{\emph{Regretted Delegation.}}
It describes cases in which the work was explicitly requested by the user, but its interpretation or execution ultimately did not match what the user wanted. Participants often located the problem in their own prompting rather than in the AI, reflecting that they should not have delegated the work in that way. In retrospect, they recognized that although they had asked the AI to perform the work, doing so had not actually reflected how they wanted the task to be handled.

\smallskip
\noindent
\textbf{\textit{Serendipitous Delegation.}}
It encompasses work that the user had neither explicitly prompted nor necessarily anticipated, but that the AI autonomously took on in a way the user found aligned with their intent or unexpectedly valuable. These cases show that autonomous initiative reveals useful possibilities the user had not considered or relieves users of work they were comfortable leaving to the AI.

\smallskip
\noindent
\textbf{\textit{Unintended Delegation.}}
It captures work that the user had not explicitly prompted and that the AI autonomously took on in a way that was not what the user wanted. Here, the AI extended the high-level delegation into work that the user did not intend it to handle in that manner.

\smallskip\noindent
\textbf{\emph{Criticality of AI-initiated work.}}
Criticality further differentiated the meaning of delegation primarily for AI-initiated work. When users had explicitly prompted the work, they almost always regarded it as important, and the few instances labeled as minor did not form a substantively distinct pattern from critical user-prompted work. In contrast, for AI-initiated work, criticality marked a meaningful difference in how participants interpreted the AI's autonomy. We therefore used criticality to further distinguish AI-initiated forms of delegation.
Within \emph{Serendipitous Delegation}, we termed critical work \textbf{\emph{Beneficial Autonomy}}, in which the AI took on work that shaped the outcome and contributed something the user valued despite not having asked for it, such as relating material the user had not thought to connect. Minor work was termed \textbf{\emph{Routine Autonomy}}, capturing low-stakes initiative that users were comfortable allowing the AI to handle on its own, such as organizing an output into a table without being asked.
% Within \emph{Serendipitous Delegation}, we termed critical AI-initiated work as \textbf{\emph{Beneficial Autonomy}}, in which the AI took on consequential work that users valued despite not having explicitly requested it, whereas minor or routine work was categorized as \textbf{\emph{Routine Autonomy}}, capturing low-stakes initiative that users were comfortable allowing the AI to handle on its own.
Within \emph{Unintended Delegation}, we termed critical work \textbf{\emph{Overreaching Delegation}}, in which the AI took on consequential work over which the user wanted to retain control or involvement. Minor work was categorized as \emph{Unnecessary Delegation}, in which the AI took on low-value or superfluous work that users viewed as unnecessary processing or token expenditure.

\subsection{Taxonomy of Desired Delegation Protocols}
\begin{table*}[t]
\centering
\small

% Vertically center the X column
\renewcommand{\tabularxcolumn}[1]{m{#1}}
\renewcommand{\arraystretch}{1.2}

\begin{tabularx}{\textwidth}{
    >{\raggedright\arraybackslash}m{0.19\textwidth}
    >{\raggedright\arraybackslash}m{0.19\textwidth}
    >{\raggedright\arraybackslash}X
    >{\centering\arraybackslash}m{0.08\textwidth}
}
\toprule

\textbf{Protocol} &
\textbf{Timing \& \newline Interaction Mode} &
\textbf{Description} &
\textbf{Autonomy} \\

\midrule

\textbf{Full Delegation} &
Silent &
The AI independently determines how to perform the work and carries it through without consulting or notifying the user. &
\textit{High} \\

\cmidrule(r){1-3}
% \hline

\textbf{Disclosed Delegation} &
Post-action notification &
The AI performs the work independently but informs the user afterward about what it did or decided, allowing the user to remain aware of and review its actions. &
$\downarrow$ \\

\cmidrule(r){1-3}
% \hline

\textbf{Calibrated Delegation} \newline
(Extent Adjustment) &
In-action adjustment \newline
(interactive) &
The AI is delegated the work, but the desired extent of its involvement is calibrated according to the user's preference---for example, by doing more, doing less, or refraining from part of the work. &
$\downarrow$ \\

\cmidrule(r){1-3}
% \hline

\textbf{Consultative Delegation} &
Pre-action consultation &
The AI consults the user before performing the work or making the relevant decision, allowing the user to confirm or redirect its course of action. &
\textit{Low} \\

\bottomrule
\end{tabularx}

\caption{Taxonomy of Desired Delegation Protocols, ordered by the relative level of autonomy afforded to the AI.}
\Description{A four-column table presenting a taxonomy of desired delegation protocols, ordered from higher to lower AI autonomy. The columns list the protocol, timing and interaction mode, description, and autonomy level. The four protocols are Full Delegation, where the AI acts silently and independently; Disclosed Delegation, where the AI acts independently but notifies the user afterward; Calibrated Delegation, where the user adjusts the extent of AI involvement during action; and Consultative Delegation, where the AI consults the user before acting.}
\label{tab:delegation-protocols}
\end{table*}
Delegation enactment describes what actually occurred, but participants' free-form comments additionally revealed how they believed the AI \emph{should} have proceeded. We refer to these preferred modes of handling delegated work as \emph{desired delegation protocols} (\autoref{tab:delegation-protocols}). Rather than simply indicating whether participants wanted more or less AI autonomy, these protocols capture where user involvement should occur relative to the AI's action and how much of the work the AI should take on.

\smallskip
\noindent
\textbf{\textit{Full Delegation.}}
This protocol describes cases where users wanted the AI to perform the work autonomously without additional consultation or notification. The relevant reasoning step could remain effectively invisible to the user as long as the AI handled it appropriately.

\smallskip
\noindent
\textbf{\textit{Disclosed Delegation.}}
This mode reflects a preference for the AI to act independently while keeping the user informed afterward about what it did or decided. The AI could retain execution authority, while the user retained visibility and an opportunity to review its actions.

\smallskip
\noindent
\textbf{\textit{Calibrated Delegation.}}
This protocol refers to cases where users were comfortable with the AI taking on the work, but wanted to adjust the extent of its involvement as the work progressed. 
Participants called for the AI to do more, do less, or refrain from particular portions of the work, indicating a need to calibrate the scope of AI involvement.

\smallskip
\noindent
\textbf{\textit{Consultative Delegation.}}
Consultation was desired when users wanted the AI to pause before proceeding with the work or committing to a consequential decision, align with them on how to continue, and obtain their input or confirmation before acting. This need for consultation did not simply arise because the user's prompt was ambiguous or required clarification. Rather, it reflected cases in which users wanted to retain authority over how the work should proceed, rather than have the AI proceed autonomously.

The protocols are ordered by the relative autonomy afforded to the AI, from \emph{Full Delegation}, where the AI can proceed independently, to \emph{Consultative Delegation}, where user input is required before proceeding. They also differ in whether and when the user is brought into the delegation process: not at all (\emph{Full Delegation}), after the AI acts (\emph{Disclosed Delegation}), as the work unfolds (\emph{Calibrated Delegation}), or before the AI acts (\emph{Consultative Delegation}). We collectively refer to the latter three as protocols that \emph{called for user involvement}.

% \smallskip
% \noindent
% The four protocols thus differ in whether the user is involved at all and, if so, when. Under \emph{Full Delegation} the AI proceeds alone, whereas the other three place user involvement at a different point relative to the AI's action: before it (\emph{Consultative}), during it (\emph{Calibrated}), or after it (\emph{Disclosed}). We refer to these three collectively as protocols that \emph{called for user involvement}. They also differ in what that involvement can change: \emph{Consultative} and \emph{Calibrated} alter the course of the work itself, while \emph{Disclosed} leaves the AI's action intact and gives the user visibility over it. We use this distinction in Section~\ref{sec:results} when the two come apart.

\subsection{Taxonomy Application and Dataset Construction}\label{sec:taxonomy-application}
We applied the finalized taxonomies to the remaining 425 data points from the 12 participants not used during taxonomy development, with four participants from each task. 
To scale the coding while assessing whether the taxonomies could be applied consistently to unseen data, we used two LLMs from different providers, GPT-5.6~\footnote{https://openai.com/index/gpt-5-6/} and Claude Opus 4.8~\footnote{https://www.anthropic.com/news/claude-opus-4-8}, as annotators. Each model received the finalized taxonomy definitions and independently assigned codes to all 425 data points. The annotation prompts are provided in the Supplementary Materials, and procedural details are reported in Appendix~\ref{app:llm-classification}.
Each model annotated \textbf{Cognitive Work} and \textbf{Desired Delegation Protocol} in separate annotation passes, with each taxonomy coded independently of the other. \textbf{Delegation Enactment} labels were derived directly from participants' structured annotations.

We assessed agreement at two levels. First, we measured agreement between the two LLM annotators across all 425 held-out data points. We then evaluated their coding against human judgments using a participant-balanced sample of 60 instances (five randomly sampled data points per held-out participant), which one author independently labeled without access to the models' predictions.
For \textbf{Desired Delegation Protocol}, the two LLM annotators agreed on 94.1\% (Cohen's $\kappa = .876$), with similarly high agreement across all annotators in the validation sample (Fleiss' $\kappa = .853$). For \textbf{Cognitive Work}, the two LLM annotators agreed on 72.7\% (Cohen's $\kappa = .686$), with a Fleiss' $\kappa$ of .656 across all annotators. These results indicate that the finalized taxonomy could be applied consistently across model annotators and showed substantial agreement with human judgments~\cite{landis1977measurement, altman1990practical}.
Disagreements between the two models were subsequently adjudicated by two authors through discussion and resolved by consensus. We then combined the resulting labels with the 567 data points used during taxonomy development, constructing a fully coded dataset of 992 reasoning steps for our empirical analyses.

% \subsection{Connecting Delegation Enactment and Desired Protocols}
% Delegation enactment and desired delegation protocols capture related but distinct aspects of the same reasoning step. The former describes how the AI actually took on and performed the work from the user's perspective, whereas the latter describes how the user believed that work should have been managed. An aligned enactment did not necessarily imply a preference for silent autonomy, nor did a misaligned enactment necessarily imply that the AI should not have performed the work at all.
% For example, \emph{Beneficial Autonomy} could correspond either to \emph{Full Delegation}, when users wanted the AI to continue taking such initiative silently, or to \emph{Disclosed Delegation}, when they welcomed the initiative but wanted to be told afterward. Likewise, \emph{Regretted Delegation} could call for \emph{Consultative Delegation}: the work was within the user's prompt, but the user wanted the AI to check its interpretation before proceeding. \emph{Overreaching Delegation} similarly called for consultation, but for a different reason---the AI had autonomously taken on consequential work the user had not intended it to decide alone. In contrast, \emph{Unnecessary Delegation} often pointed toward \emph{Calibrated Delegation}, where the issue was not whether the AI should act at all but how far its involvement should extend.

\section{Empirical Analysis}\label{sec:results}
Based on the taxonomies we built in Section~\ref{sec:taxonomy}, we analyze how delegation unfolds within LLM reasoning---what cognitive work it involves and when it extends beyond users' prompts---through three research questions.
\begin{enumerate}
    \item[\textbf{RQ1.}] What cognitive work do LLMs perform in their reasoning, and how is it organized---across tasks and within a turn?
    \item[\textbf{RQ2.}] How is delegation actually enacted inside reasoning---who initiates each step, on what cognitive work does it fall, and where does it depart from what users wanted?
    \item[\textbf{RQ3.}] How do users want delegation to proceed, and how do their preferred protocols vary across different types of cognitive work?
\end{enumerate}
Analyses cover the 992 annotated reasoning steps of Section~\ref{sec:taxonomy-application}. Because multiple reasoning steps came from the same participant, we accounted for within-participant dependence in all significance tests. We used permutation tests based on the $\chi^2$ statistic while preserving within-participant dependence. For step-level tests, labels were shuffled within participants; for position tests controlling for cognitive-work composition, labels were shuffled within participant-by-category cells; and for participant-level tests, participants were permuted across conditions.
% We used permutation tests based on the $\chi^2$ statistic, restricting permutations to preserve the relevant data structure. Step-level labels were generally shuffled within participants; when testing position effects while controlling for cognitive-work composition, labels were shuffled within participant-by-category cells. For participant-level factors such as task assignment, whole participants were permuted across conditions.

\subsection{RQ1: What Cognitive Work LLMs Perform in Reasoning}
\subsubsection{Cognitive Work Is Unevenly Distributed}
\begin{figure*}
    \centering
    \includegraphics[trim=0 0pt 0 10.5pt, clip, width=1\linewidth]{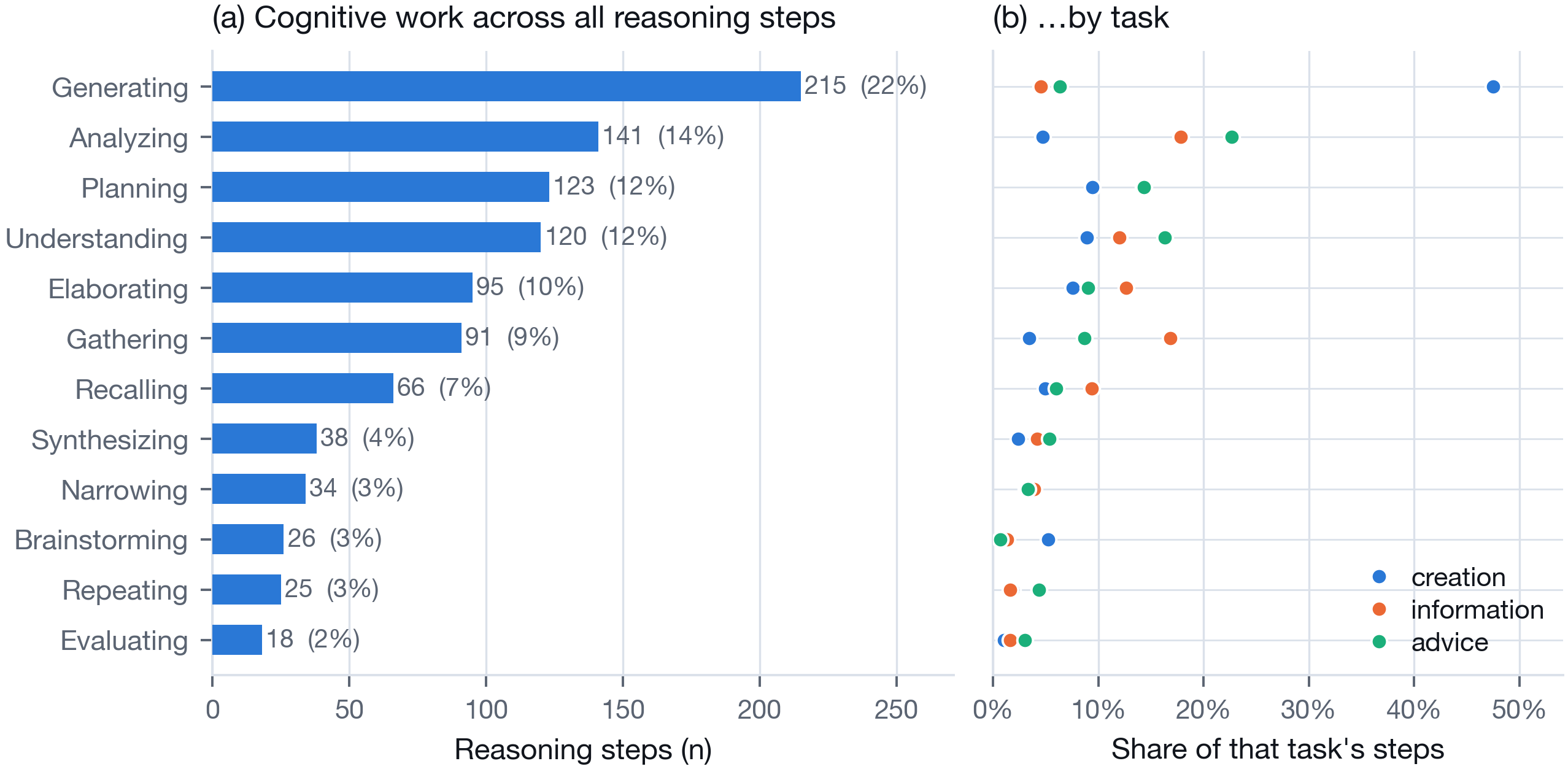}
    \caption{Cognitive work across all 992 annotated reasoning steps, by code (left) and by task category (right).}
    \Description{A two-panel chart showing the distribution of cognitive work across 992 annotated reasoning steps. Panel A is a horizontal bar chart ranking twelve cognitive-work codes by frequency: Generating is the largest with 215 steps, followed by Analyzing with 141, Planning with 123, Understanding with 120, Elaborating with 95, Gathering with 91, Recalling with 66, Synthesizing with 38, Narrowing with 34, Brainstorming with 26, Repeating with 25, and Evaluating with 18. Panel B is a dot plot showing each code's share within the Creation, Information, and Advice tasks. Generating is especially high in Creation, while Analyzing and Gathering are more prominent in Information and Advice.}
    \label{fig:rq1_cw_dist}
\end{figure*}
Cognitive work was unevenly distributed across the reasoning traces (Figure~\ref{fig:rq1_cw_dist}). \textit{Generating} was the most frequent (21.7\%), followed by \textit{Analyzing} (14.2\%), \textit{Planning} (12.4\%), and \textit{Understanding} (12.1\%).
The prominence of \textit{Generating} partly reflects how tool use surfaces in reasoning traces: each artifact-producing tool call, such as generating one slide of a pitch deck, appears as its own step, and 78\% of the 167 tool-call steps were coded \textit{Generating}. These are genuine acts of production delegated to the AI, but a single user request can fan out into many of them.

\subsubsection{Cognitive Work Differs Across Tasks}
\begin{figure*}
    \centering
    \includegraphics[width=.8\linewidth]{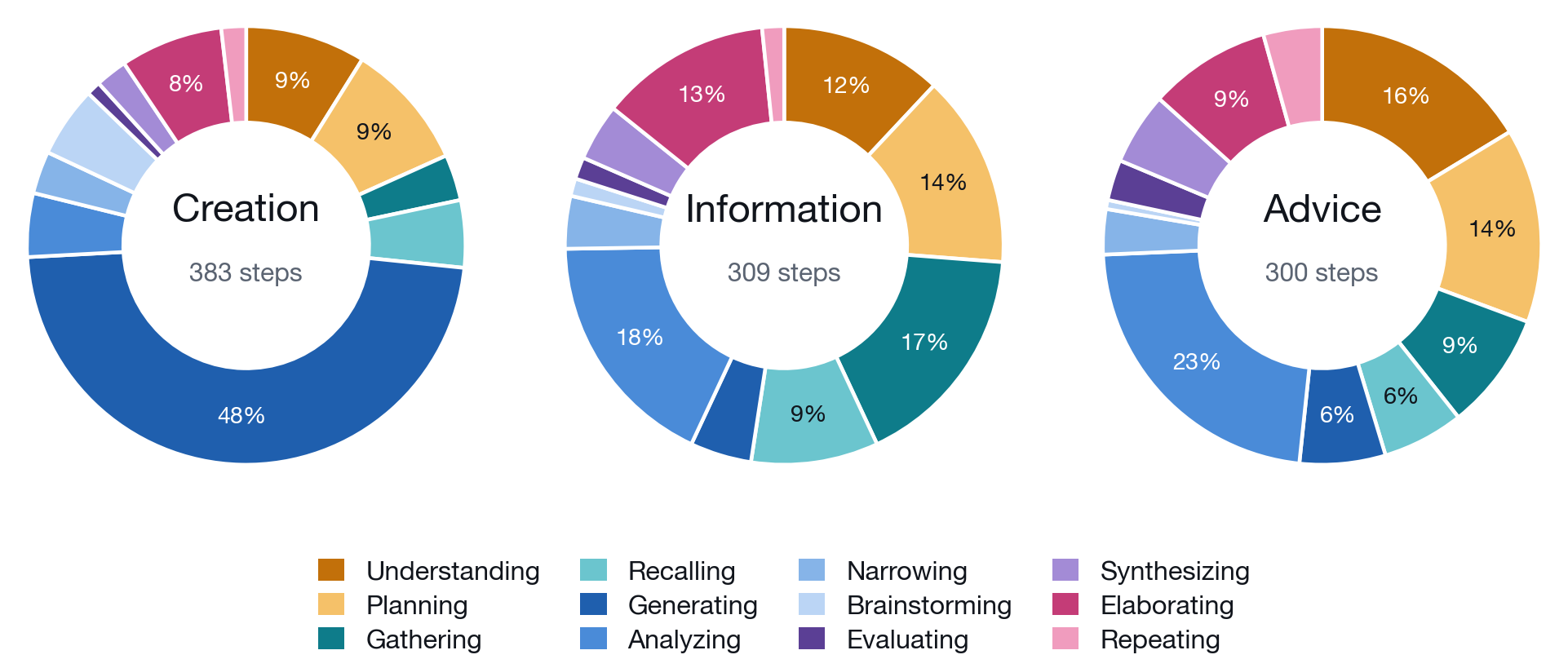}
    \caption{Composition of cognitive work within each task. Colors group the twelve codes into the five higher-level categories: orange for \emph{Framing}, teal for \emph{Sourcing}, blue for \emph{Developing}, purple for \emph{Reviewing}, and pink for \emph{Reprocessing}.}
    \Description{Three donut charts compare the composition of cognitive work within the Creation, Information, and Advice tasks. Creation has 383 steps and is dominated by Generating at 48 percent, with smaller shares for Understanding, Planning, and Elaborating. Information has 309 steps and is more evenly distributed, with Analyzing at 18 percent, Gathering at 17 percent, Planning at 14 percent, Elaborating at 13 percent, Understanding at 12 percent, and Recalling at 9 percent. Advice has 300 steps, led by Analyzing at 23 percent, followed by Understanding at 16 percent, Planning at 14 percent, Gathering and Elaborating at 9 percent each, and Generating and Recalling at 6 percent each.}
    \label{fig:rq1_cw_tasks}
\end{figure*}

The composition of cognitive work depended on the task ($\chi^2 = 320.8$, Cram\'er's $V = .402$, $p < .001$; Figure~\ref{fig:rq1_cw_tasks}). Each task had a distinct dominant code: \textit{Generating} defined Creation (47.5\% of its steps, $2.2\times$ its overall share), \textit{Gathering} was concentrated in Information ($1.8\times$), and \textit{Analyzing} dominated Advice (22.7\%, $1.6\times$). At the same time, a core of work stayed nearly constant across tasks: \textit{Understanding}, \textit{Planning}, \textit{Recalling}, \textit{Elaborating}, and \textit{Narrowing} varied by at most a few percentage points. Task-specific differences were therefore concentrated in particular forms of cognitive work, while several other types remained relatively stable across tasks.
% Tasks thus differ in the substantive work they call for, while sharing a common scaffold of framing, sourcing, and consolidating.

\subsubsection{Cognitive Work Shows Positional Tendencies Within a Turn}
\begin{figure}
    \centering
    \includegraphics[trim=0 0pt 0 10pt, clip, width=\linewidth]{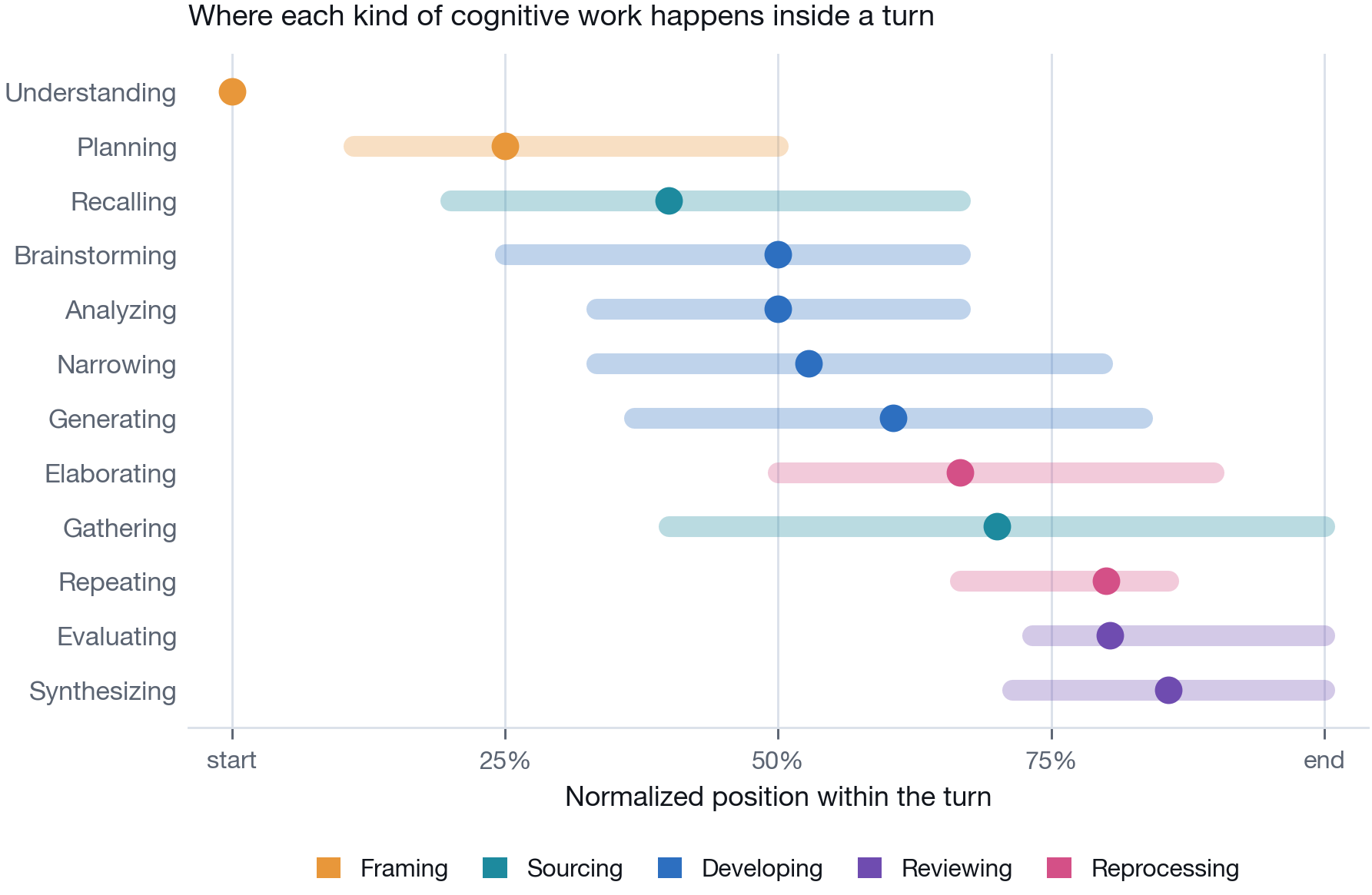}
    \caption{When each kind of cognitive work happens within a turn. Dot = median, band = interquartile range; color = higher-level category.}
    \Description{A horizontal dot-and-band chart showing where each cognitive-work code tends to occur within a turn, from start to end. Dots mark median normalized position and translucent bands mark interquartile ranges. Understanding appears at the start of turns, followed by Planning around the first quarter. Recalling, Brainstorming, Analyzing, and Narrowing cluster around the middle. Generating, Elaborating, and Gathering occur later, while Repeating, Evaluating, and Synthesizing are concentrated near the end. Colors indicate higher-level categories: Framing, Sourcing, Developing, Reviewing, and Reprocessing.}
    \label{fig:rq1_cw_position}
\end{figure}

Different types of cognitive work had different median positions within a turn, although the distributions overlapped substantially (Figure~\ref{fig:rq1_cw_position}). \textit{Understanding}, \textit{Planning}, and \textit{Recalling} had earlier median positions ($0.00$, $0.25$, and $0.40$, respectively). \textit{Generating} had a later median position ($0.61$), while \textit{Evaluating} ($0.80$) and \textit{Synthesizing} ($0.86$) had the latest medians. Overall, these suggest a progression from framing and sourcing toward developing and reviewing.

\subsection{RQ2: How Delegation Is Enacted in LLM Reasoning}
\subsubsection{Nearly Half of Steps Were Judged AI-Initiated}
\begin{figure*}
    \centering
    \includegraphics[trim=0 0pt 0 10.5pt, clip, width=1\linewidth]{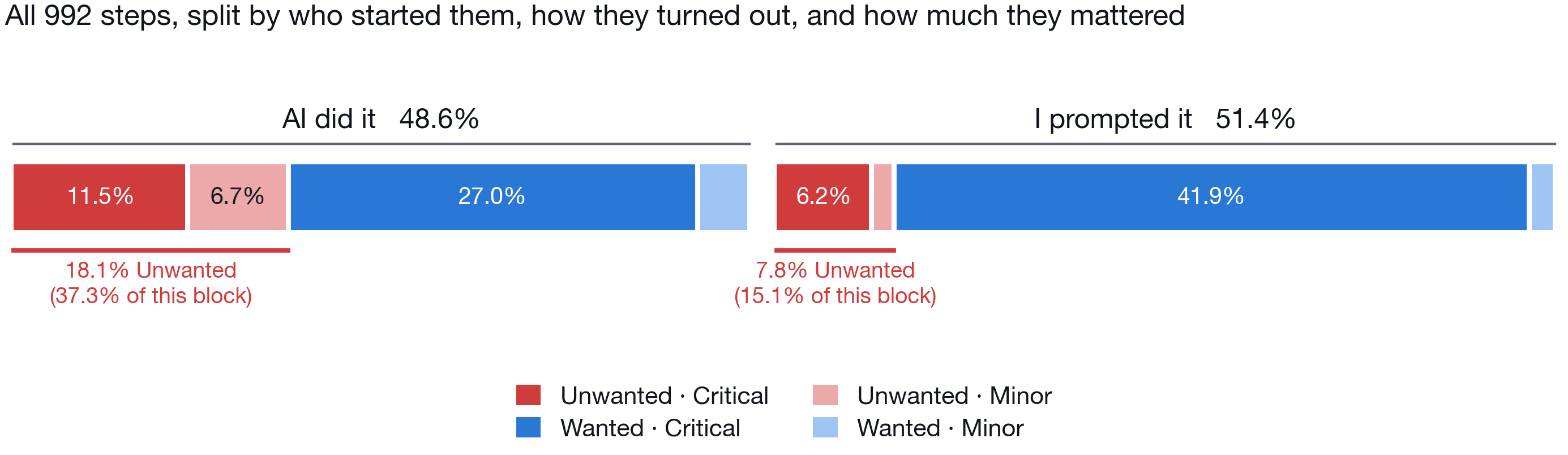}
    \caption{All 992 steps, split by who initiated them (blocks), whether their contribution was wanted (hue), and how critical they were (shade).}
    \Description{A split stacked bar chart summarizing all 992 reasoning steps by initiative, impact, and criticality. The left block shows AI-initiated steps, labeled 48.6 percent of all steps; within this block, 18.1 percent of all steps were Unwanted and 37.3 percent of AI-initiated steps were Unwanted. The largest AI-initiated segment is Wanted and Critical, at 27.0 percent of all steps. The right block shows user-initiated steps, labeled 51.4 percent of all steps; within this block, 7.8 percent of all steps were Unwanted and 15.1 percent of user-initiated steps were Unwanted. The largest user-initiated segment is Wanted and Critical, at 41.9 percent of all steps. Darker shades indicate Critical steps and lighter shades indicate Minor steps.}
    \label{fig:rq2_enactment}
\end{figure*}
Users delegated a task through their prompt, but this did not mean that they had asked the AI to take on every reasoning step involved in completing it. Participants judged 48.6\% of all steps as AI-initiated, meaning that they saw the AI as taking on work they had not requested (Figure~\ref{fig:rq2_enactment}).
Across the four forms of delegation enactment, \textsc{Intended Delegation} accounted for 43.6\% of steps, \textsc{Serendipitous Delegation} for 30.4\%, \textsc{Unintended Delegation} for 18.1\%, and \textsc{Regretted Delegation} for 7.8\%.

\subsubsection{Unwanted and Minor Steps Were More Often AI-Initiated}
AI-initiated work was not necessarily unwanted. Of the AI-initiated steps, 62.7\% contributed as users wanted (\textsc{Serendipitous Delegation}), while 37.3\% did not (\textsc{Unintended Delegation}). However, AI initiation was more common among unwanted and minor steps. 70.0\% of unwanted steps were AI-initiated, compared with 41.1\% of wanted steps, and 75.8\% of minor steps were AI-initiated, compared with 44.4\% of critical steps (both $p < .001$). 

\begin{figure}
    \centering
    \includegraphics[width=\linewidth]{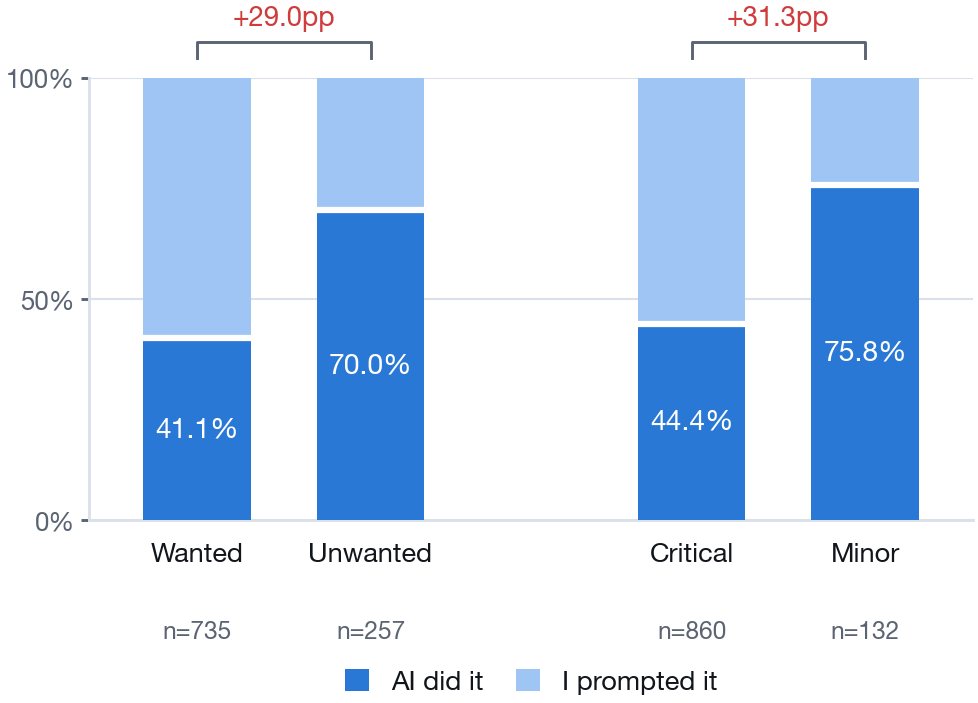}
    \caption{AI-initiated steps concentrate on work whose contribution users rated as unwanted or minor.}
    \Description{Two pairs of stacked vertical bars compare whether reasoning steps were AI-initiated or user-initiated. In the impact comparison, 41.1 percent of steps labeled Wanted were AI-initiated, compared with 70.0 percent of steps labeled Unwanted, a difference of 29.0 percentage points. In the criticality comparison, 44.4 percent of Critical steps were AI-initiated, compared with 75.8 percent of Minor steps, a difference of 31.3 percentage points. The remaining portions of each bar represent user-initiated steps.}
    \label{fig:rq2_pairs}
\end{figure}

\subsubsection{Some Kinds of Cognitive Work Invite Unwanted Initiative}
\begin{figure*}
    \centering
    \includegraphics[width=.9\linewidth]{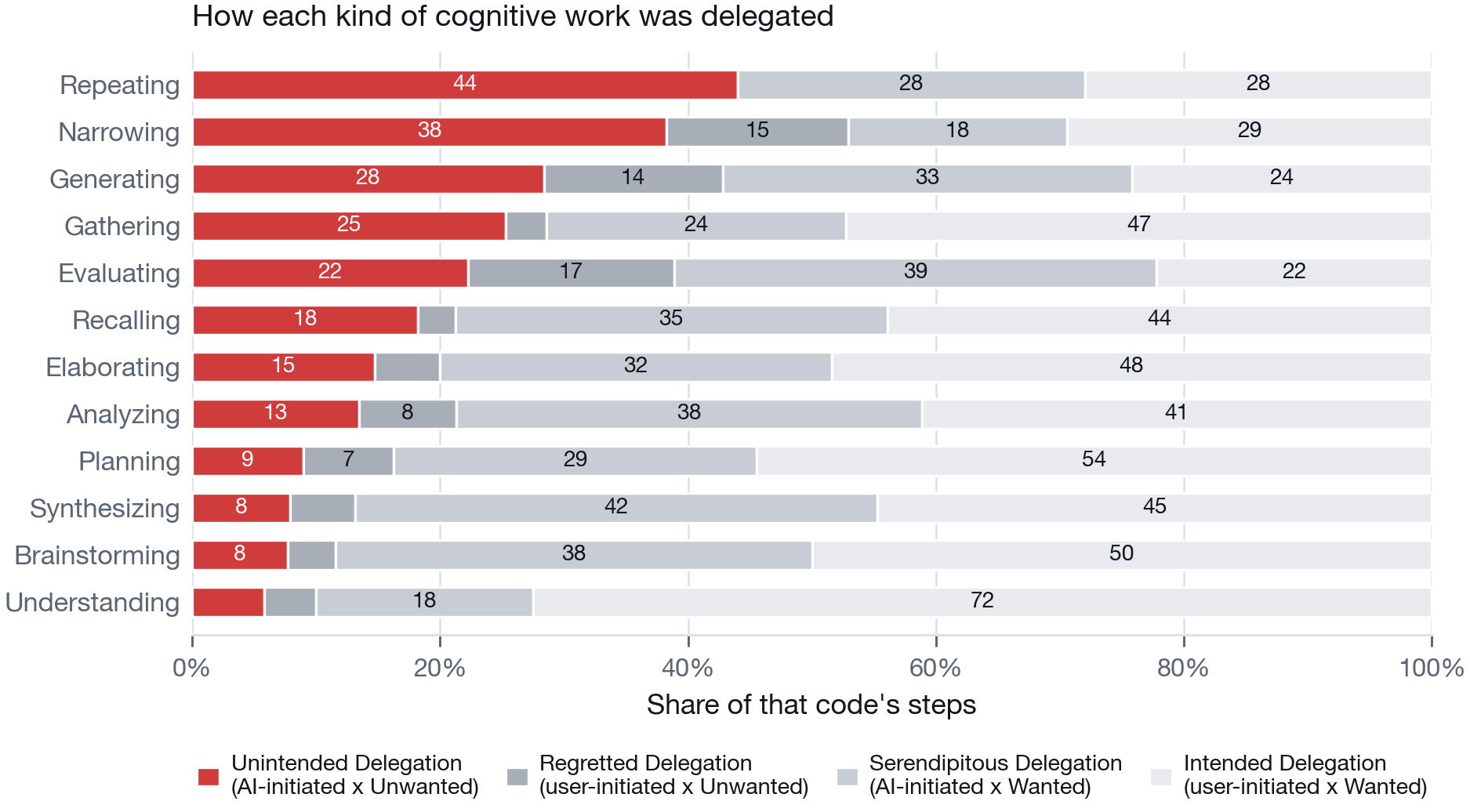}
    \caption{How each kind of cognitive work was delegated, across the four forms of delegation enactment. Codes are ordered by their share of \textsc{Unintended Delegation}.}
    \Description{A horizontal stacked bar chart showing how each cognitive-work code was delegated across four delegation enactment forms. Codes are ordered by their share of Unintended Delegation. Repeating has the highest unintended share at 44 percent, followed by Narrowing at 38 percent, Generating at 28 percent, Gathering at 25 percent, and Evaluating at 22 percent. Understanding has the lowest unintended share and the highest intended share, with 72 percent Intended Delegation. Planning also has a high Intended Delegation share at 54 percent. Serendipitous Delegation is substantial across many codes, including Elaborating, Gathering, Synthesizing, Brainstorming, Analyzing, and Recalling.}
    \label{fig:rq2_code_enactment}
\end{figure*}

Delegation enactment varied across cognitive work (Figure~\ref{fig:rq2_code_enactment}). \textit{Understanding} and \textit{Planning} were most often explicitly prompted and completed as participants wanted, with 72\% and 54\% classified as \textsc{Intended Delegation}. AI-initiated work was often welcomed in \textit{Synthesizing}, \textit{Brainstorming}, and \textit{Analyzing}, whereas \textsc{Unintended Delegation} was most common in \textit{Repeating} (44\%), \textit{Narrowing} (38\%), \textit{Generating} (28\%), and \textit{Gathering} (25\%).

These differences did not simply reflect how often the AI took initiative. \textit{Analyzing} and \textit{Gathering} had nearly identical AI-initiation rates (51\% and 50\%), but 26\% of AI-initiated \textit{Analyzing} steps were unwanted, compared with 51\% of \textit{Gathering} steps. \textit{Generating} contributed the most unintended cases by volume: 61 steps, or 34\% of all \textsc{Unintended Delegation}.

\subsubsection{Later Reasoning Steps Were More Often Judged AI-Initiated}
\begin{figure*}
    \centering
    \includegraphics[width=\linewidth]{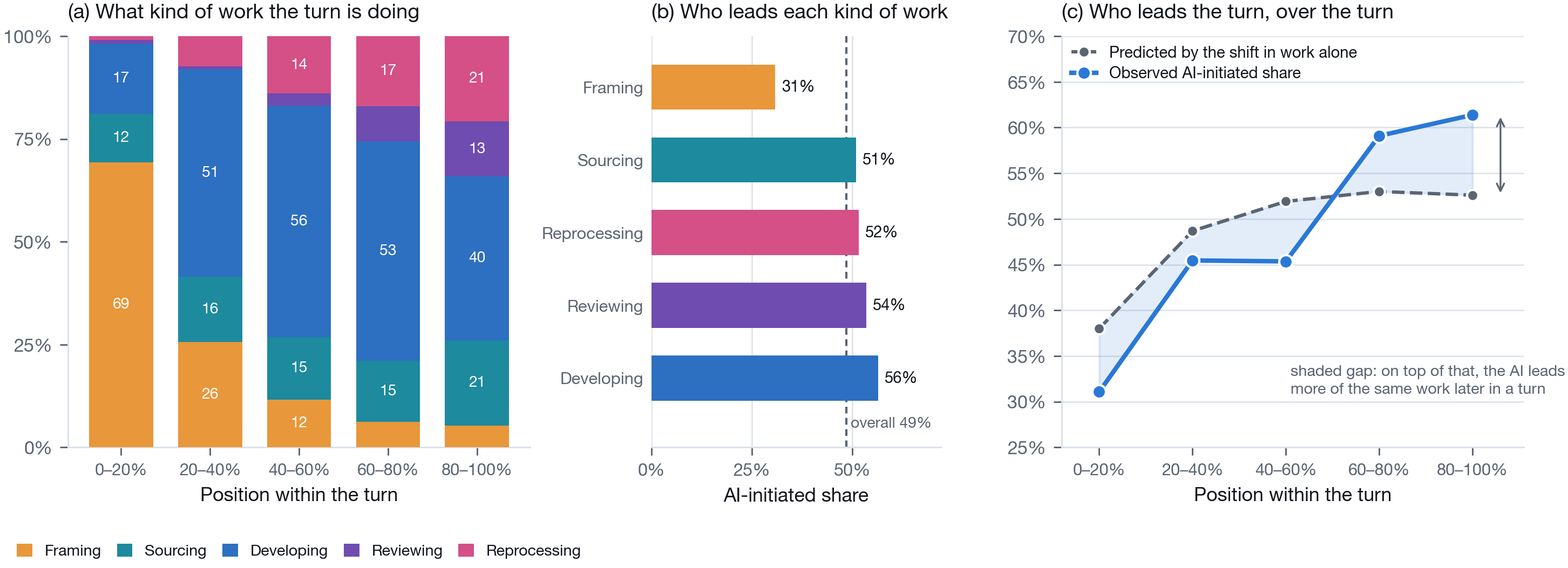}
    \caption{Across a turn's reasoning, (a) cognitive work shifts away from \textit{Framing} toward \textit{Developing}, \textit{Reviewing}, and \textit{Reprocessing}; (b) the five categories differ in how often the AI initiates them, with \textit{Framing} by far the lowest; and (c) AI initiative increases more than this compositional shift alone predicts. The dashed line in (c) applies each category's overall AI-initiation rate from (b) to the mix of cognitive work in each fifth of a turn.}
    \Description{A three-panel figure showing how cognitive work and AI initiative change over the course of a turn. Panel A is a stacked bar chart dividing turns into five position bins from 0--20 percent to 80--100 percent. Framing dominates the beginning at 69 percent but falls sharply later, while Developing is the largest category through the middle and later portions; Reviewing and Reprocessing become more common near the end. Panel B is a horizontal bar chart showing AI-initiated share by higher-level cognitive-work category: Framing is lowest at 31 percent, while Sourcing, Reprocessing, Reviewing, and Developing are around 51 to 56 percent, near or above the overall 49 percent line. Panel C is a line chart comparing observed AI-initiated share with the share predicted by changes in cognitive-work composition alone. Observed AI initiative rises from about 31 percent at the start to about 61 percent at the end, exceeding the predicted line in the last two position bins.}
    \label{fig:rq2_arc}
\end{figure*}

Participants judged 31.1\% of steps in the first fifth of a turn as AI-initiated, compared with 61.4\% in the final fifth ($p<.001$; Figure~\ref{fig:rq2_arc}). Part of this increase coincided with a shift from \textit{Framing} toward \textit{Developing}, \textit{Reviewing}, and \textit{Reprocessing}, which were more often judged AI-initiated. However, this change in cognitive-work composition predicted an increase only from 38.0\% to 52.6\%, below the observed 61.4\% (Figure~\ref{fig:rq2_arc}c). The increase therefore reflected both changing cognitive work and greater AI initiative within the same kinds of work later in the turn ($p=.008$).

\textsc{Unintended Delegation} also became more common later in the turn. However, an AI-initiated step was not significantly more likely to be unwanted later, suggesting that the increase primarily resulted from more steps being attributed to AI initiative. We report this analysis in Appendix~\ref{app:initiative-over-turn}. Reasoning spans that participants judged to be AI-initiated were also modestly farther from the immediately preceding user prompt in embedding space than those judged to be user-initiated; we report this exploratory supplementary analysis in Appendix~\ref{app:semantic-distance}.
\subsection{RQ3: How Users Want Delegation to Proceed}
For most steps (63.9\%), participants wanted the AI to proceed without consulting or notifying them. The remaining 358 steps (36.1\%) called for user involvement. Among these steps, 63\% called for consultation before the AI acted, 29\% for adjusting how much the AI did (18\% toward doing less and 11\% toward doing more), and 8\% for being informed afterward. The full distribution is shown in Appendix Figure~\ref{fig:rq3_protocol}.

\subsubsection{Different Delegation Enactments Call for Different Protocols}
\begin{figure*}
    \centering
    \includegraphics[width=0.9\linewidth]{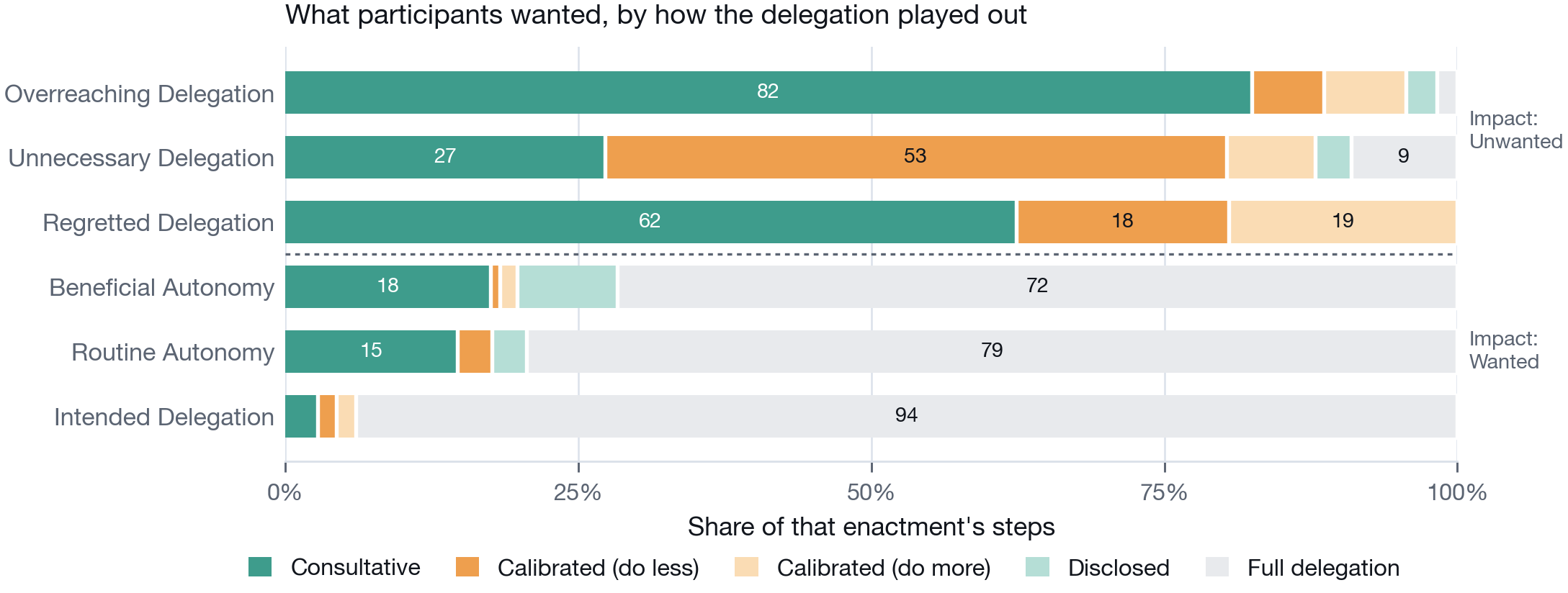}
    \caption{Desired protocols by delegation enactment, grouped by whether the step's contribution was wanted. Group sizes: \textsc{Overreaching} 114, \textsc{Unnecessary} 66, \textsc{Regretted} 77, \textsc{Beneficial Autonomy} 268, \textsc{Routine Autonomy} 34, \textsc{Intended} 433.}
    \Description{A horizontal stacked bar chart showing desired delegation protocols for six delegation enactment types, grouped by impact. For unwanted enactments, Overreaching Delegation is dominated by Consultative Delegation at 82 percent, Unnecessary Delegation is dominated by Calibrated Delegation to do less at 53 percent with 27 percent Consultative Delegation, and Regretted Delegation is mostly Consultative Delegation at 62 percent with smaller calibrated shares. For wanted enactments, Full Delegation dominates: 72 percent for Beneficial Autonomy, 79 percent for Routine Autonomy, and 94 percent for Intended Delegation. Beneficial Autonomy still has noticeable calls for involvement, including 18 percent Consultative Delegation and a small Disclosed Delegation segment.}
    \label{fig:rq3_remedies}
\end{figure*}

How delegation was enacted strongly shaped the involvement participants wanted (Cram\'er's $V=.475$, $p<.001$; Figure~\ref{fig:rq3_remedies}). For \textsc{Overreaching Delegation}---critical work the AI took on that did not contribute as participants wanted---82\% called for consultation before the AI acted. For \textsc{Unnecessary Delegation}, where the AI took on minor work participants did not want, 53\% called for the AI to do less. For \textsc{Regretted Delegation}, where participants had prompted the work but later wanted it handled differently, 62\% preferred consultation.

Successful AI initiative did not always imply full autonomy. Among \textsc{Beneficial Autonomy} steps, 28\% still called for user involvement. Moreover, 23 of the 29 requests to be informed after the AI acted occurred in these cases. Thus, participants sometimes wanted visibility into AI-initiated work even when its contribution was what they wanted.

\subsubsection{Desired Protocols Depend on the Cognitive Work}
\begin{figure*}
    \centering
    \includegraphics[width=1\linewidth]{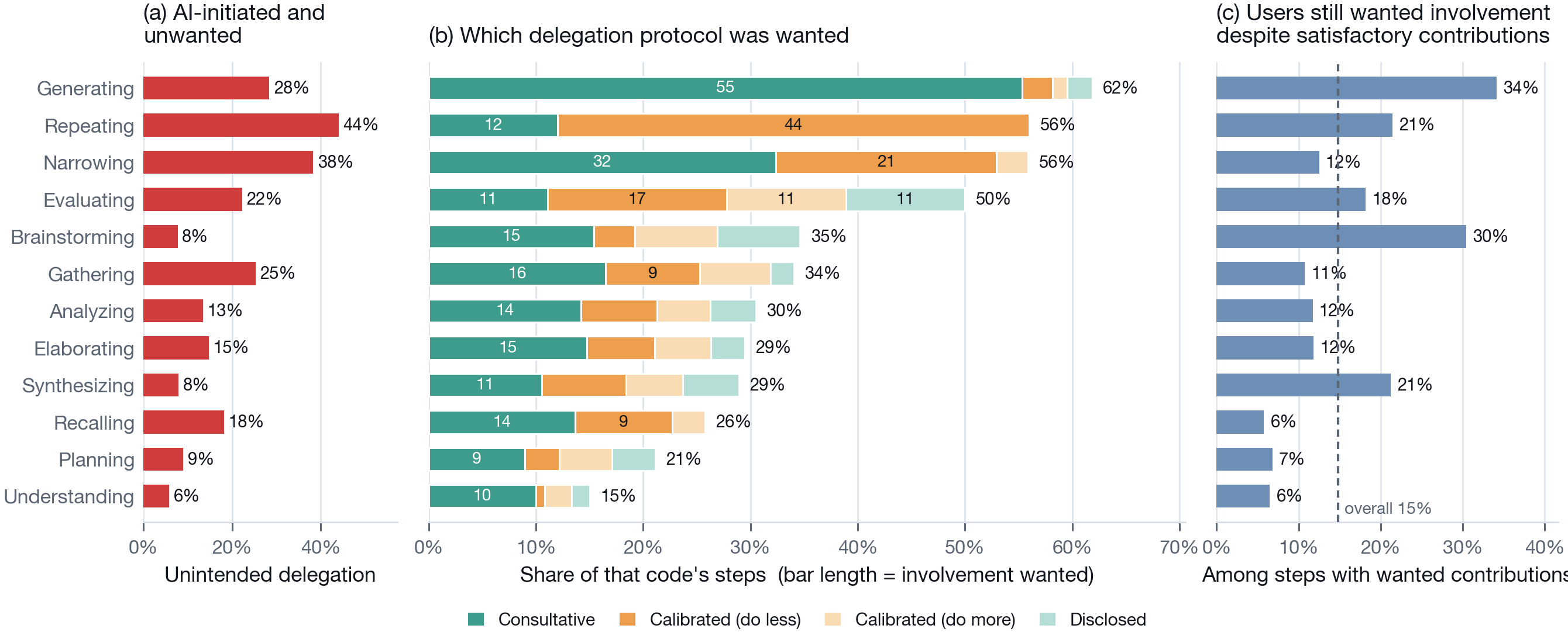}
    \caption{(a) How often each kind of cognitive work ended as \textsc{Unintended Delegation}; (b) which protocol participants wanted over it, with bar length the share of that code's steps calling for any involvement; and (c) among steps with wanted contributions, the share for which participants still wanted involvement.}
    \Description{A three-panel horizontal bar chart comparing cognitive-work codes. Panel A shows the share of each code that became Unintended Delegation: Repeating is highest at 44 percent, followed by Narrowing at 38 percent, Generating at 28 percent, Gathering at 25 percent, Evaluating at 22 percent, Recalling at 18 percent, Elaborating at 15 percent, Analyzing at 13 percent, Planning at 9 percent, Brainstorming and Synthesizing at 8 percent, and Understanding at 6 percent. Panel B shows which delegation protocol participants wanted, with total bar length indicating the share of that code's steps calling for involvement. Generating has the highest involvement demand at 62 percent, mainly Consultative Delegation; Repeating and Narrowing are both 56 percent, with Repeating dominated by Calibrated Delegation to do less and Narrowing split between consultation and calibration. Panel C shows the share for which participants still wanted involvement among steps with wanted contributions: Generating is highest at 34 percent and Brainstorming at 30 percent, both above the overall 15 percent reference line, while Understanding and Recalling are lowest at 6 percent.}
    \label{fig:rq3_code_remedies}
\end{figure*}

Desired delegation protocols also varied across cognitive work (Cram\'er's $V=.268$, $p<.001$; Figure~\ref{fig:rq3_code_remedies}). \textit{Generating} drew the greatest demand for involvement (62\%), primarily through consultation before acting (55\%). \textit{Repeating} drew a similar level of demand (56\%), but participants instead wanted the AI to do less (44\%). \textit{Narrowing} fell between these patterns, with participants wanting either consultation (32\%) or less AI involvement (21\%). In contrast, \textit{Understanding} (15\%), \textit{Planning} (21\%), and \textit{Recalling} (26\%) drew relatively little demand for involvement. 
% Similar levels of demand could therefore call for different forms of involvement depending on the work.

These differences were not simply responses to unwanted outcomes. Even among steps whose contribution matched what participants wanted, the desire for involvement continued to vary across cognitive work (Figure~\ref{fig:rq3_code_remedies}c). Among satisfactory steps, only 6\% of \textit{Recalling} and \textit{Understanding} called for involvement, compared with 34\% of \textit{Generating} and 30\% of \textit{Brainstorming}. Notably, only 8\% of \textit{Brainstorming} resulted in unwanted delegation, yet participants still wanted involvement in nearly a third of its satisfactory steps. 
% Users therefore wanted a role in some kinds of work regardless of whether the AI's contribution was satisfactory.

These patterns suggest that involvement served different purposes across cognitive work. Participants wanted to shape work that determined substantive content or constrained the solution space, as in \textit{Generating}, \textit{Brainstorming}, and \textit{Narrowing}. For \textit{Repeating}, involvement instead meant limiting unnecessary work. Participants were more willing to leave preparatory or background work, such as \textit{Understanding}, \textit{Planning}, and \textit{Recalling}, to the AI. Desired involvement therefore reflected the work at stake, not only whether the AI's contribution was satisfactory.

\section{Design Exploration: Toward Deliberate Delegation}\label{sec:design}
Our analysis shows that a single task request can encompass multiple kinds of cognitive work that users may want delegated in different ways. The three taxonomies in Section~\ref{sec:taxonomy} help translate this finding into design by identifying what work the AI performed, how it was delegated, and how users wanted the AI to proceed.
Drawing on the taxonomies and empirical findings, we develop two design implications for supporting more deliberate delegation---enabling users to decide more selectively what work the AI takes on and when and how they remain involved. We illustrate these implications with five low-fidelity design sketches that make concrete how interfaces could represent delegated work and let users shape how it proceeds.

\subsection{DI1: Support Multiple Delegation Protocols Within a Single Response}
A single LLM response involves multiple kinds of cognitive work, and users did not want to delegate all of it the same way: users wanted consultation before some AI-initiated generation, calibration for repetitive work, disclosure of work they wanted to see, and full delegation for work they were comfortable leaving to the AI. These patterns can indicate where coordination may be needed, but should not prescribe fixed protocols. Building on these patterns, an interface could translate the relevant cognitive work into concrete task decisions and let the user choose how the AI should proceed. We explore how this involvement could be configured before the task, connected to the current request while prompting, and enacted during and after execution.

\begin{figure*}[t]
    \centering
    \includegraphics[width=\linewidth]{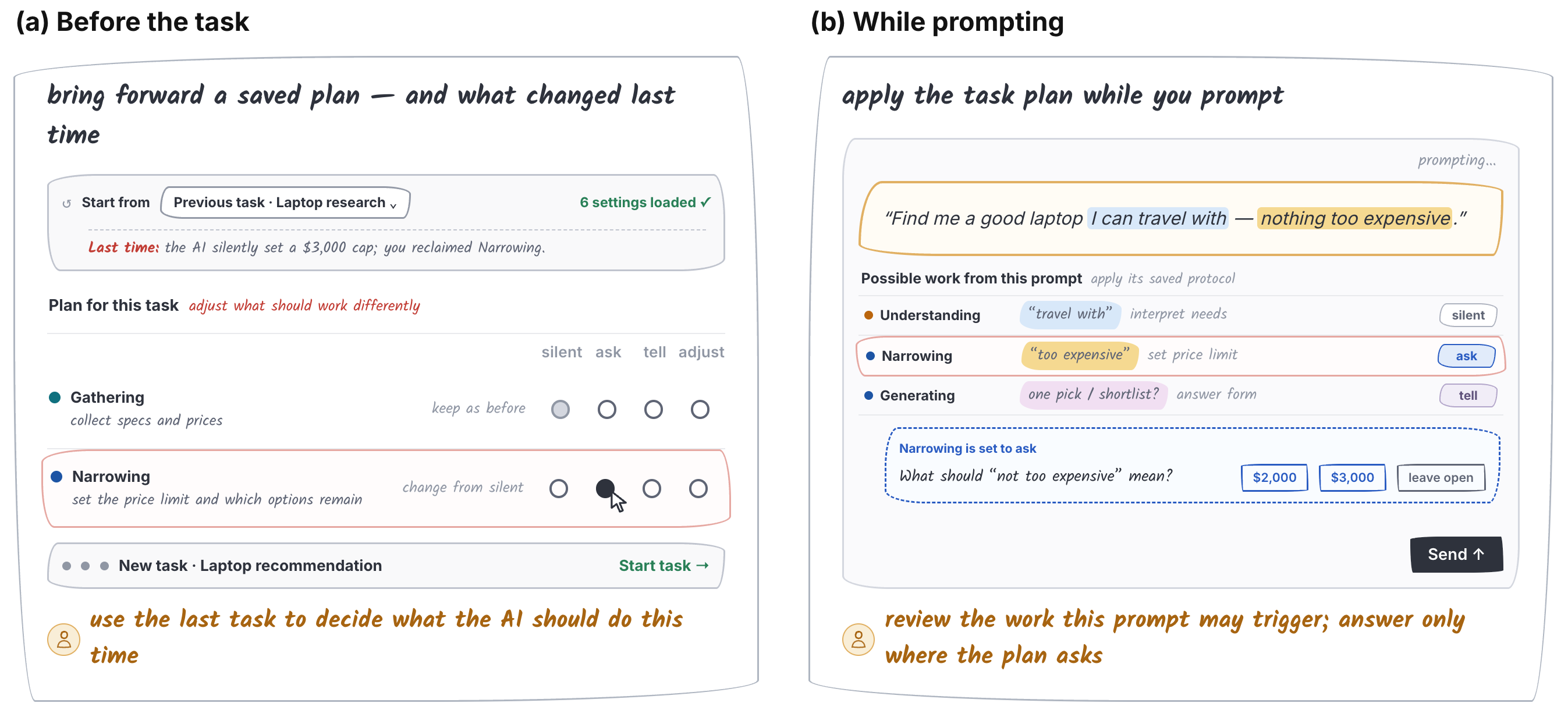}
    \caption{Configuring delegation before work begins. (a) A plan reuses protocols explicitly selected in a previous task and foregrounds the decisions the user may want to handle differently. (b) During prompt composition, possible cognitive work is linked to prompt wording and the saved plan; a decision assigned to Ask opens a request for user input.}
    \Description{Two interface panels for a laptop recommendation task. In panel a, a previous-task configuration loads six settings and recalls that the user reclaimed Narrowing after an inferred 3,000-dollar cap. Two rows show Gathering set to Silent and Narrowing set to Ask. A cursor points to Ask; Start task appears at the bottom. In panel b, the draft prompt highlights travel with and nothing too expensive. Rows link Understanding to interpreting travel needs with Silent, Narrowing to setting a price limit with Ask, and Generating to an answer format with Tell. Below them, a price question offers 2,000 dollars, 3,000 dollars, or Leave open, followed by Send.}
    \label{fig:di1-1}
\end{figure*}

\paragraph{\textbf{Before the task.}}
Users could review a delegation plan specifying which cognitive work the AI should carry out independently and where they want to be involved (Figure~\ref{fig:di1-1}a). The plan could start from protocols the user selected in previous tasks and show a past episode to help them decide what to change. For laptop selection, \textit{Gathering} specifications and prices might remain \textit{Silent}, while a previous incident in which the AI applied an unwanted price cap leads the user to set \textit{Narrowing} to \textit{Ask}. This gives users a preview of how the AI will carry out the task and lets them decide which work to delegate or handle together.

\paragraph{\textbf{While prompting.}}
As users compose a prompt, the interface could show what cognitive work it may lead the AI to perform and how the delegation plan would apply (Figure~\ref{fig:di1-1}b). In ``find me a good laptop I can travel with---nothing too expensive,'' \textit{Understanding} concerns interpreting travel needs, \textit{Narrowing} concerns applying a price limit, and \textit{Generating} concerns producing the recommendation. Because \textit{Narrowing} is set to \textit{Ask}, the interface opens a question about the price limit. This could help users notice, before sending the prompt, that the AI may otherwise choose a budget on their behalf. They can then set that budget or leave it to the AI, making their intended division of work clearer without specifying every intermediate step.

\begin{figure*}[t]
    \centering
    \includegraphics[width=\linewidth]{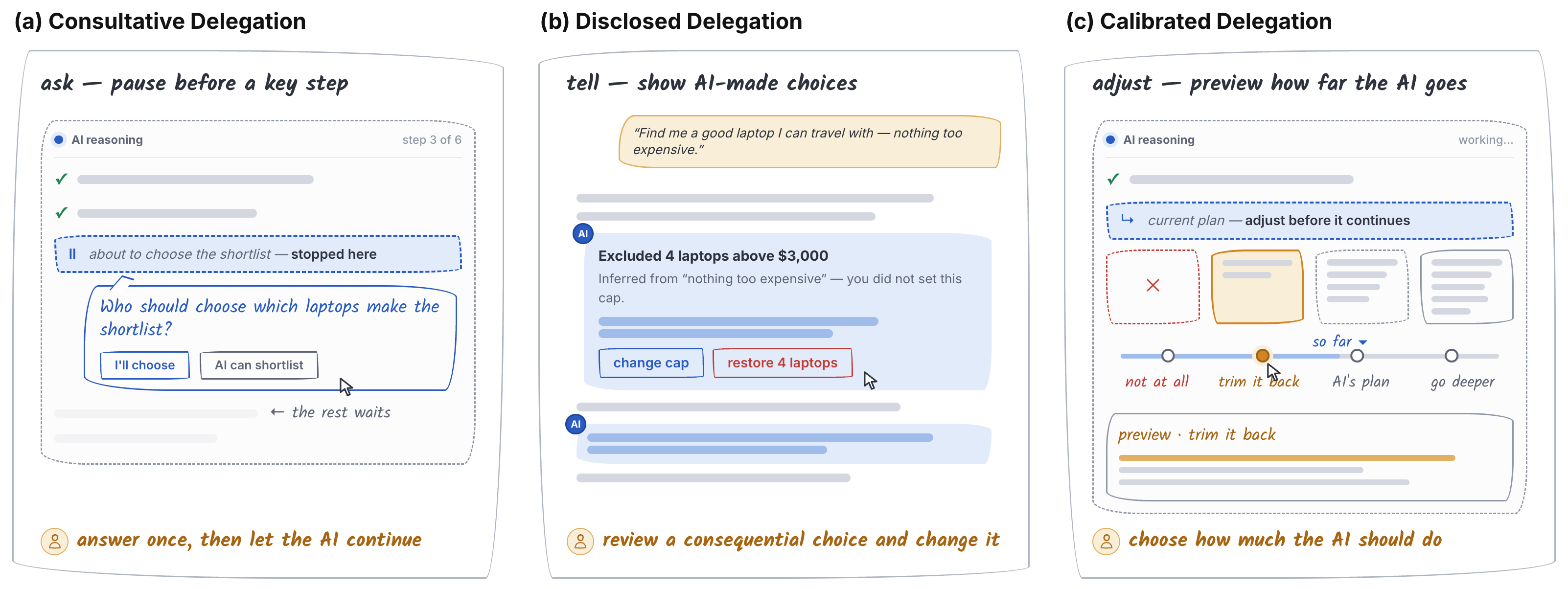}
    \caption{Three ways to involve the user within one response. (a) Ask pauses before a selection decision. (b) Tell discloses a consequential AI-initiated choice within the answer and offers ways to revise it. (c) Adjust previews different extents of ongoing work. These interaction forms instantiate Consultative, Disclosed, and Calibrated Delegation, respectively; Full Delegation proceeds without an additional interaction.}
    \Description{Three panels arranged left to right. In Ask, an artificial intelligence reasoning panel pauses at shortlist selection. A speech bubble asks who should choose the laptops, with I'll choose and AI can shortlist buttons; faded rows below indicate waiting work. In Tell, a user prompt sits above an answer containing two blue highlighted passages with AI badges. The larger passage states that four laptops above 3,000 dollars were excluded under a cap inferred from nothing too expensive, with Change cap and Restore 4 laptops buttons. In Adjust, a reasoning panel contains four previews and a slider labeled Not at all, Trim it back, AI's plan, and Go deeper. Trim it back is selected, and a preview appears below. Cursors indicate actionable controls in all three panels.}
    \label{fig:di1-2}
\end{figure*}

\paragraph{\textbf{During and after execution.}}
The interface could let users ask the AI to pause, review a choice it has made, or adjust work in progress (Figure~\ref{fig:di1-2}). Under \textsc{Consultative Delegation}, \textit{Ask} pauses before shortlist selection and lets the user decide whether they or the AI should choose the laptops. Under \textsc{Disclosed Delegation}, \textit{Tell} identifies a choice within the answer: the AI interpreted ``nothing too expensive'' as a \$3,000 cap, excluding four laptops. Showing the criterion and the affected options could help users decide how to steer the next interaction. They could change the cap or restore the excluded laptops without having to reconstruct why those options disappeared. Under \textsc{Calibrated Delegation}, \textit{Adjust} previews different extents of ongoing work, allowing users to ask the AI to do less, stop, or go further. These sketches show how users could apply different delegation protocols within a single response: leaving some work to the AI, retaining a role in selected decisions, and revising its work afterward. Users could thereby participate where they want input or control while allowing other work to proceed independently.

\subsection{DI2: Make Fine-Grained Delegation Visible as Interaction Unfolds}
Delegation continues to take shape after the user sends a prompt, as the AI takes on different kinds of cognitive work while producing a response. Our findings also showed that users sometimes wanted visibility into AI-initiated work even when its contribution was wanted. Interfaces could therefore make it easier to see what work the AI took on, how it approached that work, and where the user may want to intervene. Such visibility could support deliberate delegation by helping users connect the AI's actions to their own intentions and decide how to proceed together.

\begin{figure*}[t]
    \centering
    \includegraphics[width=0.95\linewidth]{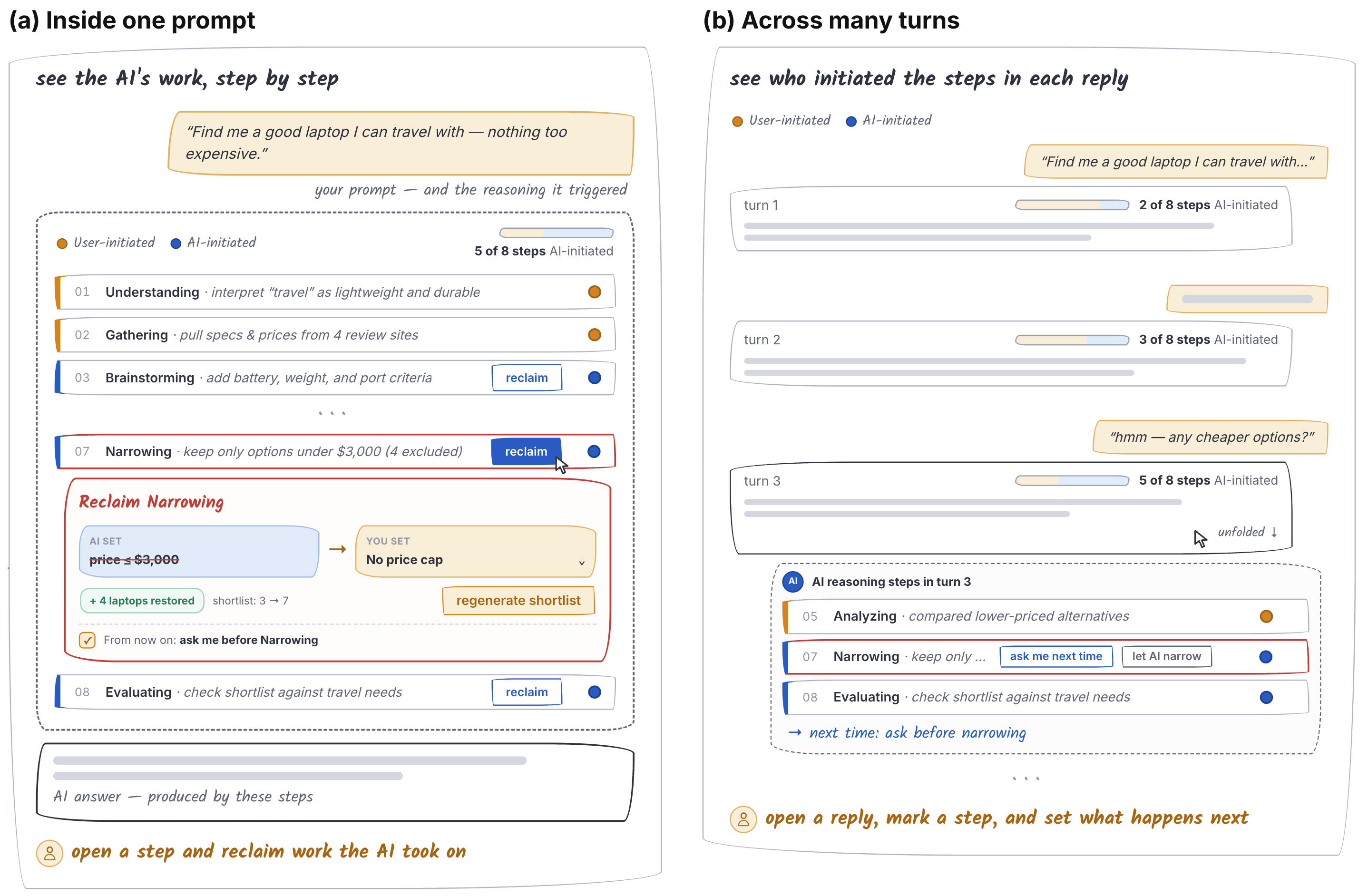}
    \caption{Inspecting and revising delegation within and across responses. (a) Concrete reasoning steps carry cognitive-work and initiative labels. Opening Reclaim on \textit{Narrowing} exposes the inferred criterion, the effect of changing it, and controls to regenerate the shortlist and request future consultation. (b) A compact summary inside each reply shows user- and AI-initiated step counts relative to that reply's total reasoning steps. Expanding a reply reveals individual steps and controls for future delegation. The bars count attributed steps, not the amount of thinking contributed; counts are illustrative.}
    \Description{Two panels show a travel-laptop conversation. Panel (a) shows reasoning steps between the prompt and answer, with amber and blue markers for user- and AI-initiated steps. A summary indicates that 5 of 8 steps were AI-initiated. AI-initiated steps can be reclaimed; the example removes an AI-set \$3,000 price cap, restores four laptops, and lets the user choose whether the AI should ask before narrowing in the future. Panel (b) shows the interaction across turns, with each reply summarizing AI-initiated steps using explicit counts such as 2 of 8, 3 of 8, and 5 of 8. Turn 3 is expanded to show step-level initiative and future control options.}
    \label{fig:di2-1}
\end{figure*}

\begin{figure*}[t]
    \centering
    \includegraphics[width=\linewidth]{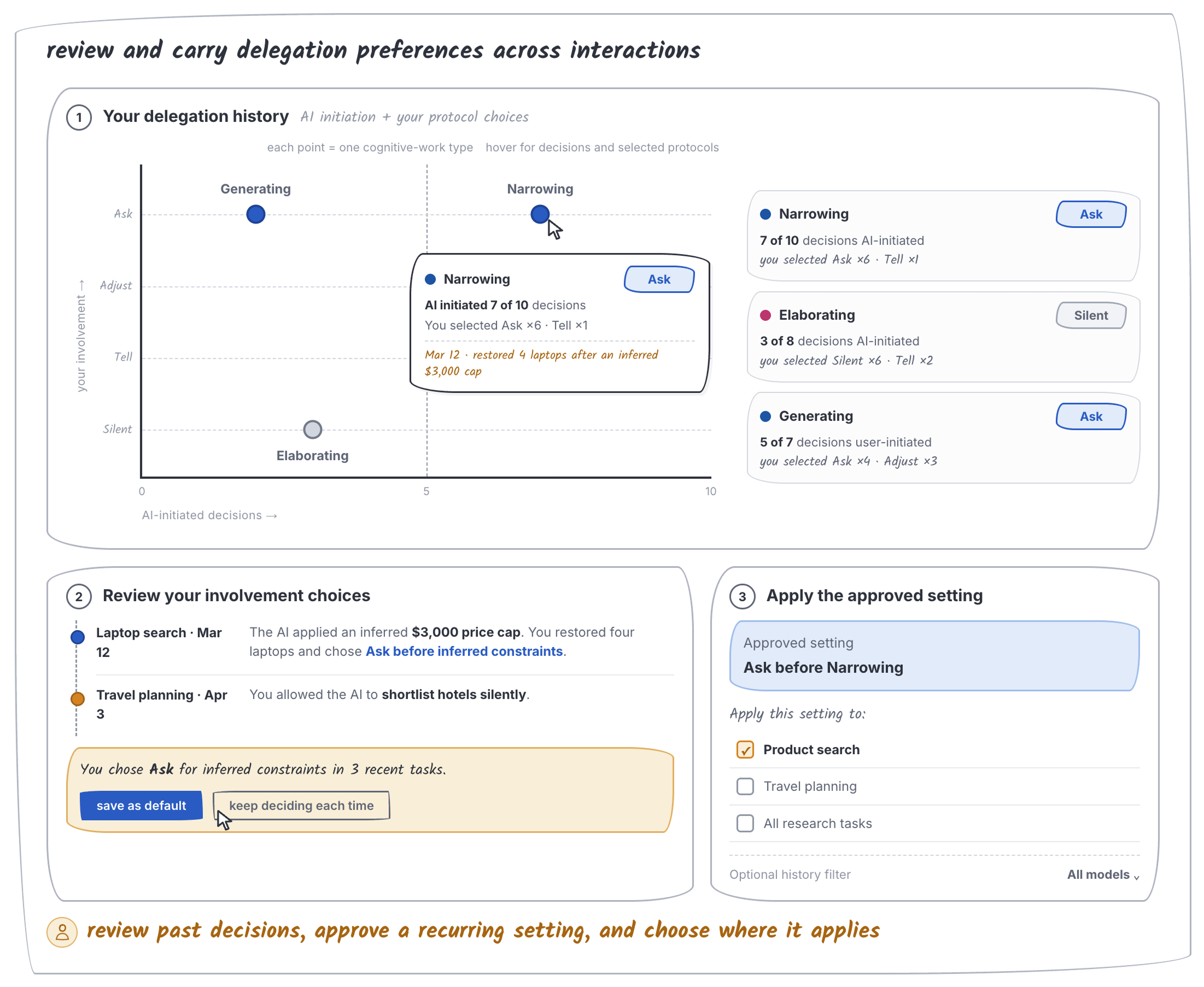}
    \caption{Reviewing delegation across interactions through an illustrative personal history. A cognitive-work overview connects recorded initiative and selected protocols to specific past decisions. Users can review recurring choices, approve a default, and select the task types to which it applies. Counts and episodes are design examples, not longitudinal study results.}
    \Description{A dashboard with one wide top panel and two panels below. The top graph has AI-initiated decision counts from zero to ten on the horizontal axis and involvement settings Silent, Tell, Adjust, and Ask from bottom to top. Generating is at two and Ask, Narrowing at seven and Ask, and Elaborating at three and Silent. A cursor on Narrowing opens a card with seven of ten decisions AI-initiated, six Ask selections, one Tell selection, and an episode restoring four laptops after an inferred 3,000-dollar cap. Summary cards beside the graph repeat Narrowing and show Elaborating with three of eight decisions AI-initiated and six Silent and two Tell selections; Generating has five of seven decisions user-initiated and four Ask and three Adjust selections. The lower-left panel lists a March 12 laptop decision and an April 3 hotel-shortlisting decision. It offers Save as default or Keep deciding each time after three Ask choices for inferred constraints. The lower-right panel shows Ask before Narrowing as the approved setting. Product search is checked, while Travel planning and All research tasks are unchecked. An optional history filter is set to All models.}
    \label{fig:di2-2}
\end{figure*}

\paragraph{\textbf{Within a response.}}
Users could inspect reasoning steps labeled with their cognitive-work type and attributed initiative (Figure~\ref{fig:di2-1}a). A summary reports how many steps were attributed to AI initiative, while the underlying steps remain accessible. In the example, opening ``Reclaim'' on a \textit{Narrowing} step reveals that the AI applied a \$3,000 price cap and excluded four laptops. Users can change the criterion, regenerate the shortlist, or request consultation before similar decisions. The same interaction could support involvement without rejecting the result: users might keep a satisfactory shortlist while reviewing its selection rule or requesting consultation next time.

\paragraph{\textbf{Across turns.}}
A conversation overview could show how many of the attributed steps in each reply were AI-initiated (e.g., 5 of 8), while retaining access to the underlying work (Figure~\ref{fig:di2-1}b). Opening a reply would reveal its cognitive work, allowing users to inspect a decision and choose how similar work should proceed later without reviewing every step or restarting the task.

% A conversation overview could summarize user- and AI-initiated steps within each reply while retaining access to the underlying work (Figure~\ref{fig:di2-1}b). These summaries would represent counts of attributed steps, not the amount of thinking contributed by either party. Opening a reply reveals the cognitive work it contains, allowing users to inspect a decision and choose how similar work should proceed in later turns. For instance, after reviewing the price cap applied during \textit{Narrowing}, a user could select \textit{Ask me next time} or \textit{Let AI narrow}. This could help users revisit how a conversation reached its current state and redirect the AI at the level of a particular decision. They would not need to review every step or restart the task to change how one kind of work is delegated.

\paragraph{\textbf{Across tasks.}}
A personal dashboard could help users review how they have delegated different kinds of cognitive work across interactions (Figure~\ref{fig:di2-2}). Its overview places each work type by the number of AI-initiated decisions and the user's current involvement setting. Inspecting a point reveals the underlying counts, protocol selections, and a past episode. The review panel could help users consider those choices in context: they may have asked for consultation after an inferred laptop price cap while allowing the AI to shortlist hotels silently in another task. When a choice recurs, the interface could offer to save it as a default and let the user select the task types to which it applies. This could help users carry useful ways of working with the AI into future tasks while retaining the ability to choose differently in other contexts.

Overall, these sketches show how interfaces could support deliberate delegation: users could choose how different cognitive work should proceed, inspect and revise work the AI takes on, and carry those choices into later interactions.
% The same history could also be organized by cognitive-work type. Selecting Narrowing, for example, could bring together price filtering, hotel shortlisting, and literature selection from different chats, with links to the original prompts and decisions. This would give users a way to find and compare how the AI handled similar work without searching through entire conversations. Together, the overview and access to past episodes could support reflection on what users want to delegate, where they want a role, and which arrangements they wish to reuse. These views extend our findings into a design possibility for longer-term use, with future settings grounded in choices users review and approve.
\section{Discussion}
We discuss applications of the three taxonomies to reasoning analysis and post-training, adaptive AI proactivity, and deliberate delegation, followed by the limitations of this work.

\subsection{Taxonomies for Reasoning Analysis and Post-Training}
Our taxonomies may help structure process-level preference data for post-training by capturing three aspects of delegation: what cognitive work the AI performs, how it takes on that work, and how the user wants to be involved. Existing process supervision uses intermediate correctness labels to train reasoning reward models and refine agent trajectories~\cite{lightman2024verify,xiong2024watch}. Our taxonomies could complement these correctness signals with labels for delegation and user involvement. At scale, such data could support evaluators of beneficial and unwanted initiative or preference models that learn when an AI should act autonomously and how it should coordinate with the user.

Building such training data requires analyzing how work unfolds, not only task outcomes. While AgentBoard tracks milestones in agent trajectories~\cite{ma2024agentboard} and Evalet decomposes outputs into functional fragments~\cite{kim2026evalet}, our taxonomies add the user's perspective on delegation. Applied across models, prompts, tools, or agent architectures, they could reveal where AI initiative exceeds a request, whether users welcome that initiative, and how systems with similar outputs divide cognitive work differently between AI and user.

Our dataset could be small and task-specific to support such post-training directly; instead, the taxonomies suggest a schema for collecting process-level preferences at scale. These preferences should remain conditioned on the user and context rather than become general rules. Moreover, the framework analyzes inspectable model-produced traces and tool actions, not internal cognition, which chain-of-thought explanations may not faithfully represent~\cite{turpin2023unfaithful}.

\subsection{Taxonomies for Adaptive AI Proactivity}
Our taxonomies may support adaptive AI proactivity by framing proactivity as when and how to involve users, not merely how often an AI acts without being asked. This distinction matters because proactive support can assist but also disrupt users~\cite{pu2025assistance}; situational autonomy can outperform fixed autonomy~\cite{salikutluk2024evaluation}; and user checkpoints can correct errors while adding cognitive load~\cite{he2025plan}.

The three taxonomies provide complementary inputs: Cognitive Work identifies the work underway, Delegation Enactment captures how the AI took it on, and Desired Delegation Protocol indicates how the user wanted to participate. These inputs could help a system decide whether to proceed silently, ask before acting, allow adjustment, or disclose an action afterward. Whereas prior work identifies when human intervention would help~\cite{feng2024rehac}, our framework extends this decision to what form that involvement should take.

Rather than prescribe fixed mappings, group-level patterns could inform correctable defaults. DI1 suggests supporting different protocols for different cognitive work within a response, while DI2 suggests making AI-initiated work visible and carrying reviewed choices into later interactions. As users respond to these interventions, systems could adapt when to proceed silently and when to ask, disclose, or support adjustment. Mixed-initiative dialogue offers one mechanism for learning from such corrections~\cite{arakawa2025scaling}.

\subsection{From Task-Level Collaboration to Deliberate Delegation}
We argue for \emph{deliberate delegation} in human--AI collaboration: making allocations of cognitive work visible and revisable without requiring user intervention at every step. Existing accounts often describe collaboration through tasks, workflow stages, interaction modes, or conversational turns~\cite{gao2024taxonomy,holter2024deconstructing,song2026good}. Yet a single response may contain multiple kinds of cognitive work, including work the user expected to retain. A step-level view reveals how this work is allocated and where users may want different forms of involvement.

DI1 treats involvement as a choice among coordination protocols, rather than a level of automation. Users may prefer guided participation to full automation for exploratory or creative work~\cite{song2026good,khurana2025do}. By supporting consultation, adjustment, disclosure, or full delegation for different cognitive work, interfaces could preserve useful AI initiative without requiring involvement at every step.

DI2 concerns what an interface should make visible and revisable. Decomposing AI work can support steering and verification~\cite{kazemitabaar2024steering}, while provenance can help users retain control and ownership~\cite{hoque2024hallmark}. Rather than expose every step, interfaces could surface AI-initiated work that may warrant attention, let users inspect and revise specific decisions, and carry reviewed choices into later interactions.
Deliberate delegation thus combines selective involvement with visibility into how cognitive work is allocated, allowing users to retain a role without giving up useful AI initiative.

\subsection{Limitations and Future Work}
Our findings come from 24 frequent LLM users and three predefined tasks. Each of the Creation, Information, and Advice categories was represented by one task, so differences between categories may also reflect differences between the specific tasks. The tasks were short and carried fewer consequences and organizational constraints than many forms of knowledge work. Participants also used one model, one interface, and a limited set of tools. Model behavior, system prompts, trace format, and available tools can all change which steps appear. For example, the PowerPoint tool represented separate artifact-producing calls as separate steps, increasing the share of \textit{Generating} in the Creation task. Future work should apply the taxonomies across populations, domains, tasks, models, and agent architectures before treating the distributions reported here as general patterns.

Participants evaluated each step retrospectively after seeing the output. This context helped them connect intermediate work to its effects, but it may also have introduced hindsight and outcome biases. Reviewing every step after a task also differs from being interrupted during ongoing work. Stated preferences for consultation, adjustment, or disclosure may change when those mechanisms impose real costs. The design sketches in Section~\ref{sec:design} are therefore proposals rather than evaluated interfaces. Future studies should implement these protocols and measure task performance, interruption, workload, control, ownership, and the retention of useful AI initiative over time.

The taxonomies also simplify work that can serve more than one function. We assigned one primary cognitive-work code to each step, and agreement for Cognitive Work was lower than for Desired Delegation Protocol. The categories may need to be extended or represented hierarchically in specialized domains. Further human validation and cross-domain use could clarify their boundaries. The post-training and adaptive-policy directions also require larger preference datasets and tests of whether learned policies generalize across contexts. Such systems should avoid treating group-level patterns as user-specific preferences and consider the privacy and agency implications of maintaining a history of what each user chooses to delegate.

\section{Conclusion}
We examined how users perceived delegation within LLM reasoning traces. Drawing on 992 retrospective step-level annotations from 24 participants, we developed three taxonomies characterizing what cognitive work each reasoning step performed, how that work was delegated, and how users wanted the AI to proceed. Participants viewed nearly half of the steps as AI-initiated, with this share increasing as reasoning progressed. Desired involvement varied across cognitive work and delegation enactment, including when contributions were wanted. These findings highlight the importance of understanding delegation at the reasoning-step level, where users may want different roles across the work involved in a single task. Our design exploration illustrates how this understanding could support deliberate delegation by helping users inspect AI-initiated work and choose when and how to participate.

\bibliographystyle{ACM-Reference-Format}
\bibliography{references}

\clearpage
\appendix

\section{LLM-Based Taxonomy Application Procedure}
\label{app:llm-classification}

After developing the taxonomies through qualitative analysis, we used LLMs to code the held-out portion of the dataset for two finalized taxonomies: \textbf{Cognitive Work} and \textbf{Desired Delegation Protocol}. 
This appendix provides the detailed procedure for this LLM-assisted coding process, including input construction, prompting, reliability assessment, and adjudication. 
The full prompt templates used for LLM annotation are provided in the supplementary materials: one for Cognitive Work classification and one for Desired Delegation Protocol classification.

\subsection{Annotation Scope}

The unit of annotation was an individual reasoning step or tool-use action produced by the LLM during a participant's task session. 
Although labels were assigned at the step level, each step was classified in context rather than as an isolated text span. 
This context was important because several labels depended on relationships among steps. 
For example, \emph{Elaborating} and \emph{Repeating} required identifying whether a step returned to content from a prior reasoning step, while \emph{Synthesizing} required recognizing when the model integrated multiple earlier reasoning steps into a consolidated conclusion.
We also provided preceding turns from the same session when they were needed to interpret the target turn. 
This was especially important for \emph{Recalling}, which required determining whether the model was bringing forward information from earlier conversation history.

We used LLM annotators for two taxonomies: \textbf{Cognitive Work} and \textbf{Desired Delegation Protocol}. These were classified in separate annotation passes using separate prompts. \textbf{Delegation Enactment} was not classified by the LLM annotators; it was derived directly from participants' structured annotations of initiative, criticality, and impact.

\subsection{Input Construction}
For each target turn, we constructed an annotation packet containing the interaction context needed to classify its reasoning steps. 
Each packet included the user's prompt, the AI's final response, the ordered reasoning steps and tool-use actions in that turn, and the participant's retrospective annotation and free-form explanation for each step. 
The packet also included preceding turns from the same session to provide conversational context. 
The prompt included the finalized taxonomy definitions and coding rules for the taxonomy being applied.

For Cognitive Work classification, each LLM annotator selected the type of work performed by each reasoning step, such as \emph{Understanding}, \emph{Planning}, \emph{Gathering}, \emph{Generating}, or \emph{Elaborating}. 
For Desired Delegation Protocol classification, each LLM annotator selected how the participant wanted the AI to proceed with that work, such as \emph{Full Delegation}, \emph{Disclosed Delegation}, \emph{Calibrated Delegation}, or \emph{Consultative Delegation}. 
When classifying Desired Delegation Protocol, the participant's free-form explanation was treated as the primary evidence because the taxonomy captures the participant's preferred form of involvement.

\subsection{Prompting Procedure}
We prompted two LLMs from different providers, GPT-5.6-sol and Claude-opus-4.8, as independent annotators. 
The two models received the same annotation packets and taxonomy definitions, but their labels were generated independently. 
We used one prompt template for Cognitive Work classification and a separate prompt template for Desired Delegation Protocol classification.

In the Cognitive Work pass, the model was instructed to assign exactly one cognitive-work code to each reasoning step and to provide a brief rationale grounded in the supplied context. 
For steps labeled as \emph{Elaborating} or \emph{Repeating}, the model was additionally instructed to identify which prior reasoning step the current step elaborated on or repeated. 
This allowed us to preserve the relational structure of \emph{Reprocessing}, whose meaning depends on a step's connection to earlier reasoning.

In the Desired Delegation Protocol pass, the model was instructed to assign exactly one desired-protocol label to each reasoning step and to provide a brief rationale grounded in the participant's retrospective explanation and the surrounding turn context. 
Because this taxonomy captures how participants wanted the AI to proceed, the model was instructed to prioritize participants' free-form explanations when they clarified the desired form of involvement.

\subsection{Reliability Assessment and Adjudication}
We assessed the consistency of LLM-based taxonomy application in two ways. 
First, we measured agreement between the two LLM annotators across the 425 held-out data points. 
For Cognitive Work, they agreed on 72.7\% of labels, with Cohen's $\kappa = .686$.
For Desired Delegation Protocol, the two models agreed on 94.1\% of labels, with Cohen's $\kappa = .876$.

Second, we compared the model labels against human judgments using a participant-balanced validation sample of 60 instances, consisting of five randomly sampled data points from each held-out participant. 
One author independently labeled this sample without access to the models' predictions. 
Agreement across annotators was substantial for Cognitive Work, with Fleiss' $\kappa = .656$, and high for Desired Delegation Protocol, with Fleiss' $\kappa = .853$.

After reliability assessment, disagreements between the two LLM annotators were adjudicated by two authors through discussion and resolved by consensus. 
The adjudicated labels for the 425 held-out data points were then combined with the 567 data points used during taxonomy development, resulting in a fully coded dataset of 992 reasoning steps for the analyses reported in Section~\ref{sec:results}.

% -----------------
\section{Additional Analyses of AI Initiative Across a Turn}
\label{app:initiative-over-turn}
The cognitive-work composition changed substantially across a turn. \textit{Framing} declined from 69.4\% of steps in the first fifth to 5.3\% in the final fifth, while \textit{Developing}, \textit{Reviewing}, and \textit{Reprocessing} together increased from 18.9\% to 74.0\% ($\chi^2=405.1$, Cram\'er's $V=.320$, $p<.001$). \textit{Framing}, the work most directly tied to the user's prompt, was judged AI-initiated in 30.9\% of its steps, compared with 51--57\% for the other four higher-level categories ($p<.001$).

\begin{figure*}
    \centering
    \includegraphics[width=\linewidth]{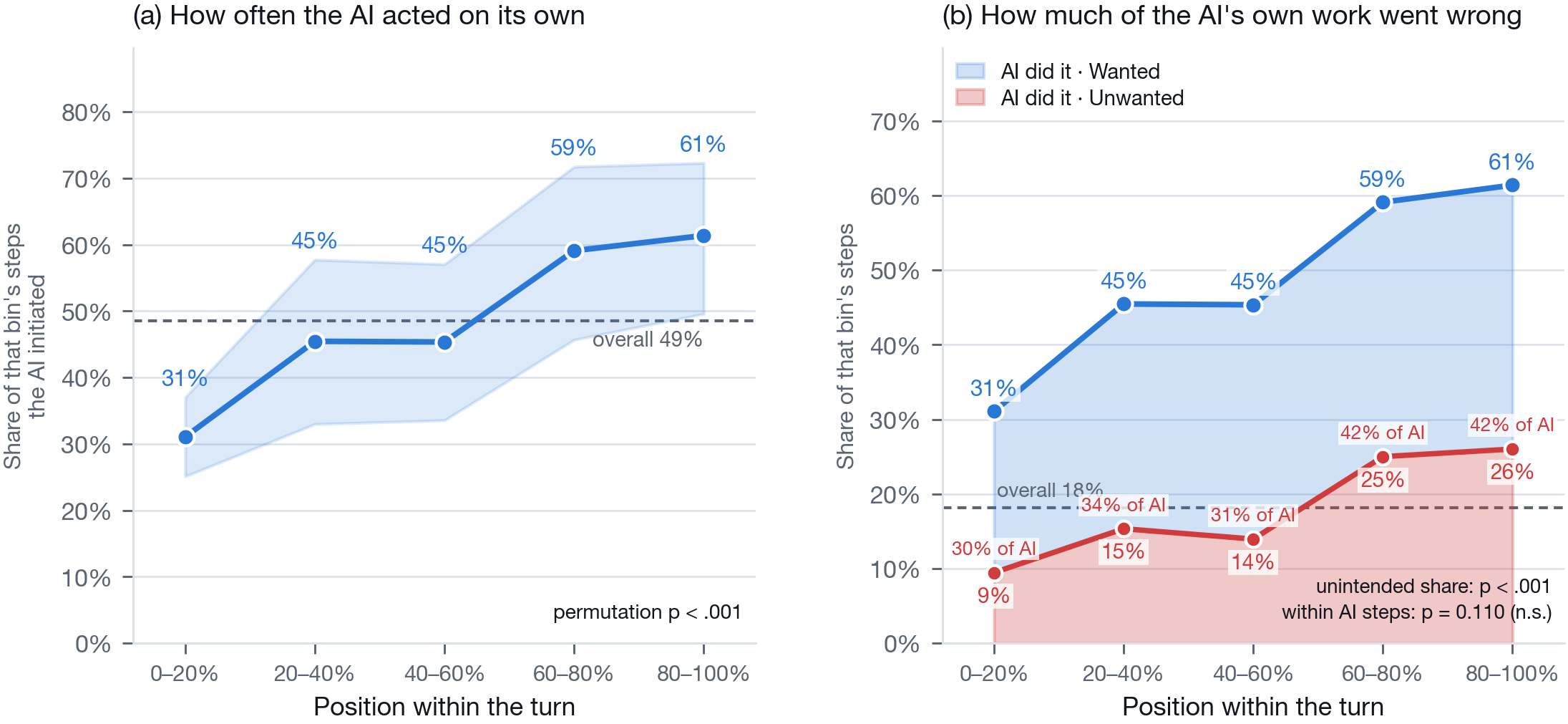}
    \caption{Across a turn's reasoning, (a) the share of steps judged AI-initiated and (b) the share classified as \textsc{Unintended Delegation}. Bands indicate 95\% participant-level cluster bootstrap intervals.}
    \Description{A two-panel line chart showing changes across five normalized positions within a turn. Panel A shows the share of steps judged AI-initiated, which increases toward the end of the turn. Panel B shows that the share of Unintended Delegation also increases, while the proportion of AI-initiated steps that were unwanted does not change significantly.}
    \label{fig:rq2_over_turn}
\end{figure*}

The increase in AI initiative was accompanied by more \textsc{Unintended Delegation}, which rose from 9.5\% of all steps in the first fifth to 26.0\% in the final fifth ($2.8\times$, $p<.001$; Figure~\ref{fig:rq2_over_turn}). However, among AI-initiated steps, the share that was unwanted did not increase significantly (30\% to 42\%, $p=.110$). The increase in unintended delegation therefore resulted primarily from participants attributing more steps to AI initiative, rather than from each AI-initiated step becoming more likely to be unwanted.

\section{Prompt Similarity and Perceived Initiative}
\label{app:semantic-distance}
As an exploratory supplementary analysis, we examined whether participants' initiative judgments were associated with the semantic similarity between an annotated reasoning span and the user prompt for the corresponding turn. This analysis tests for a group-level association; it does not establish causality or test whether similarity can classify individual reasoning spans.

\paragraph{Measure.}
For each annotation $a$, let $t_a$ be the reasoning text the participant selected and $p_a$ the immediately preceding user prompt. Writing $\mathbf{e}(\cdot)$ for the embedding of a passage, we compute their cosine similarity, $s(a)$:
\begin{equation}
  s(a) \;=\; \frac{\mathbf{e}(t_a) \cdot \mathbf{e}(p_a)}
                  {\lVert \mathbf{e}(t_a) \rVert \; \lVert \mathbf{e}(p_a) \rVert}
  \label{eq:rq3-sim}
\end{equation}
Higher values indicate greater similarity, or shorter semantic distance from the prompt. We embedded each prompt and reasoning span as a single passage using \texttt{gemini-embedding-001} with the \texttt{SEMANTIC\_SIMILARITY} task setting and 768 dimensions. The unit is the participant's original annotation: a single judgment over one contiguous stretch of reasoning. When an annotation covered multiple reasoning steps, we concatenated the selected segments in order so that the embedded text matched the scope of that judgment. The analysis includes original reasoning-span annotations from all 24 participants.

\begin{figure*}[t]
  \centering
  \begin{subfigure}[b]{0.58\textwidth}
    \centering
    \includegraphics[width=\linewidth]{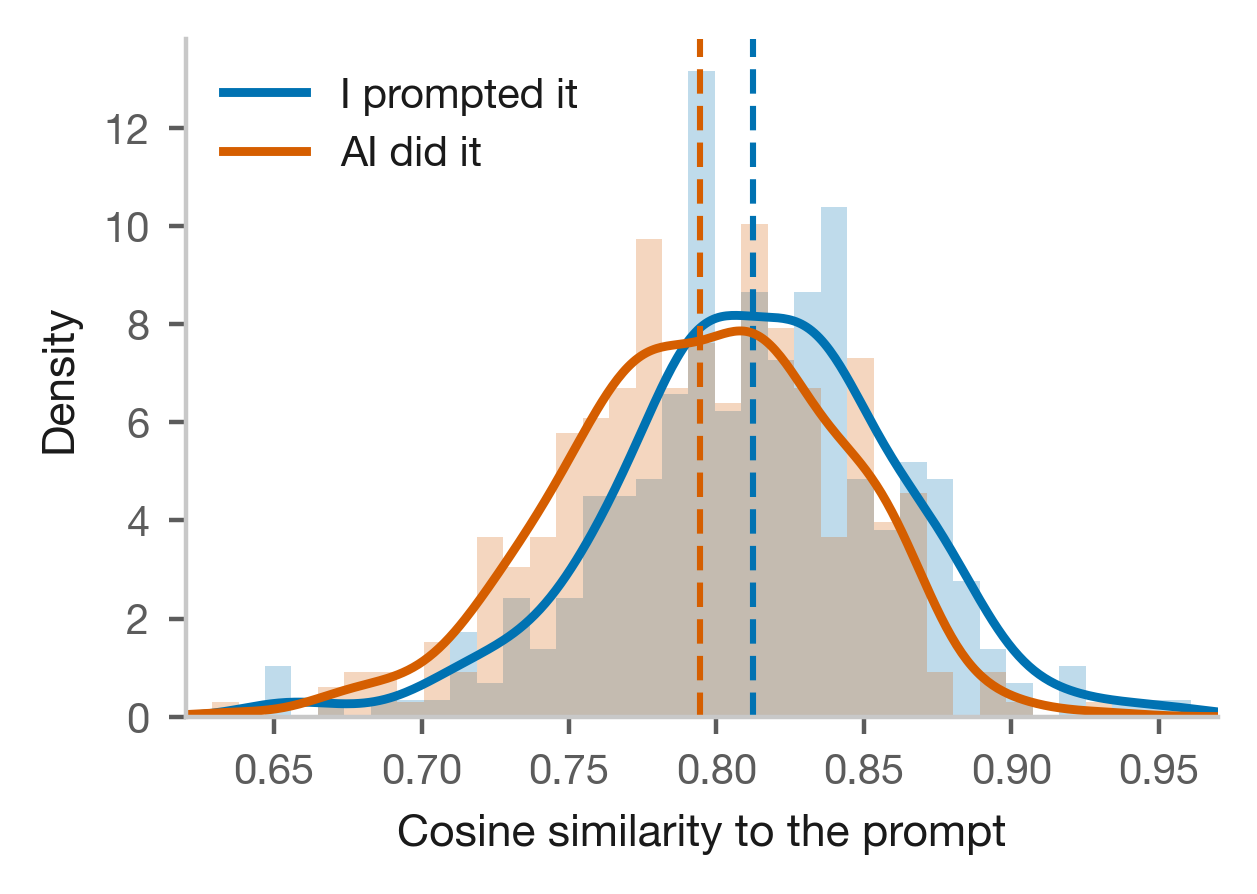}
    \caption{}
    \label{fig:rq3-distribution}
  \end{subfigure}
  \hfill
  \begin{subfigure}[b]{0.4\textwidth}
    \centering
    \includegraphics[width=\linewidth]{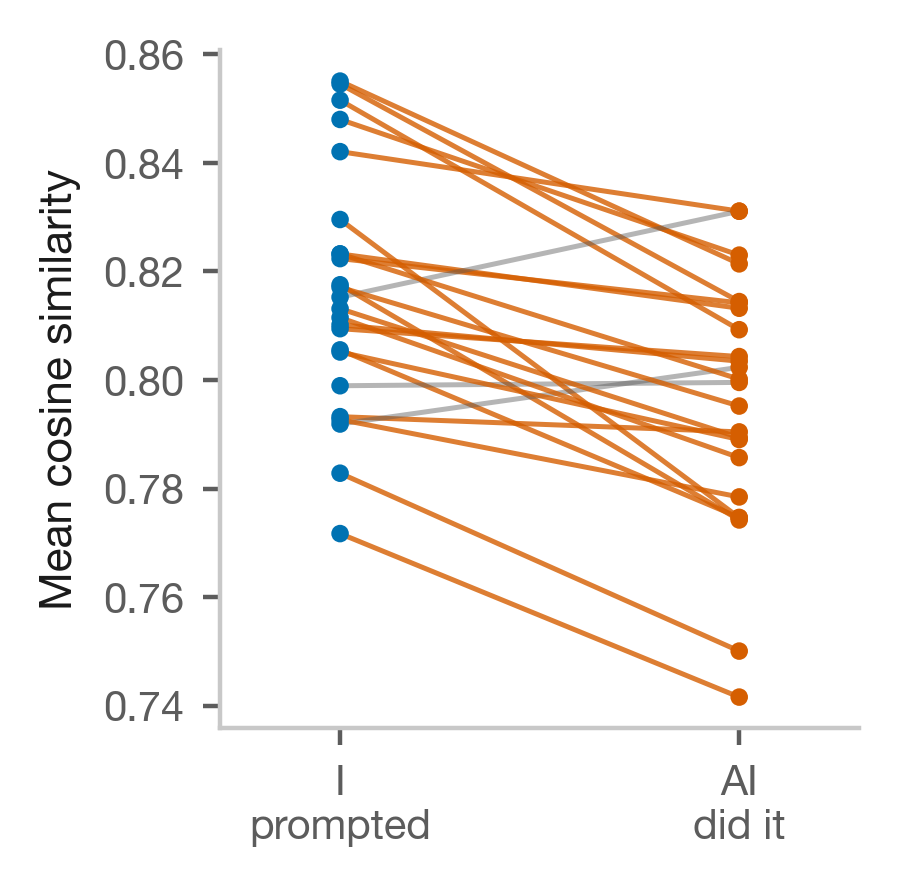}
    \caption{}
    \label{fig:rq3-participants}
  \end{subfigure}
  \caption{Prompt similarity by perceived initiative. (a) Distribution over the reasoning-span annotations; dashed lines mark the group means. The two distributions overlap substantially; their mean difference is $-0.018$, equivalent in magnitude to $0.37$ pooled standard deviations. (b) Each line joins one participant's two group means. Twenty-one of the 24 lines slope downward.}
  \Description{Left: two overlapping density curves, the one for reasoning attributed to the AI sitting slightly left of the one for reasoning the participant prompted. Right: a slope chart with 24 lines, most descending from left to right.}
  \label{fig:rq3-1}
\end{figure*}

\paragraph{Result.}
AI-initiated annotations had a slightly lower mean similarity of $0.795$ ($SD = 0.046$), compared with $0.813$ ($SD = 0.048$) for user-initiated annotations. The mean difference was $-0.018$, equivalent in magnitude to $0.37$ pooled standard deviations, with Cliff's $\delta = -0.217$. The two distributions nevertheless overlapped substantially (Figure~\ref{fig:rq3-distribution}).

The direction and approximate size of the difference were similar across the three task conditions (creation $\delta = -0.199$, information $-0.256$, advice $-0.219$). Twenty-one of the 24 participants also had a lower mean for AI-initiated annotations (Figure~\ref{fig:rq3-participants}). A permutation test that reshuffled initiative labels within participants yielded $p < .0001$ ($20{,}000$ draws). Together, these results indicate a modest aggregate association, not a reliable separation of individual annotations.

Overall, prompt similarity showed a modest group-level association with participants' initiative judgments. This correlational analysis does not establish why participants made those judgments. It also compares each reasoning span only with the immediately preceding prompt, so work licensed by earlier conversational context may appear more distant. Given the substantial overlap between the distributions, this result alone does not establish that prompt similarity can reliably identify individual AI-initiated annotations.

\begin{figure*}
    \centering
    \includegraphics[width=1\linewidth]{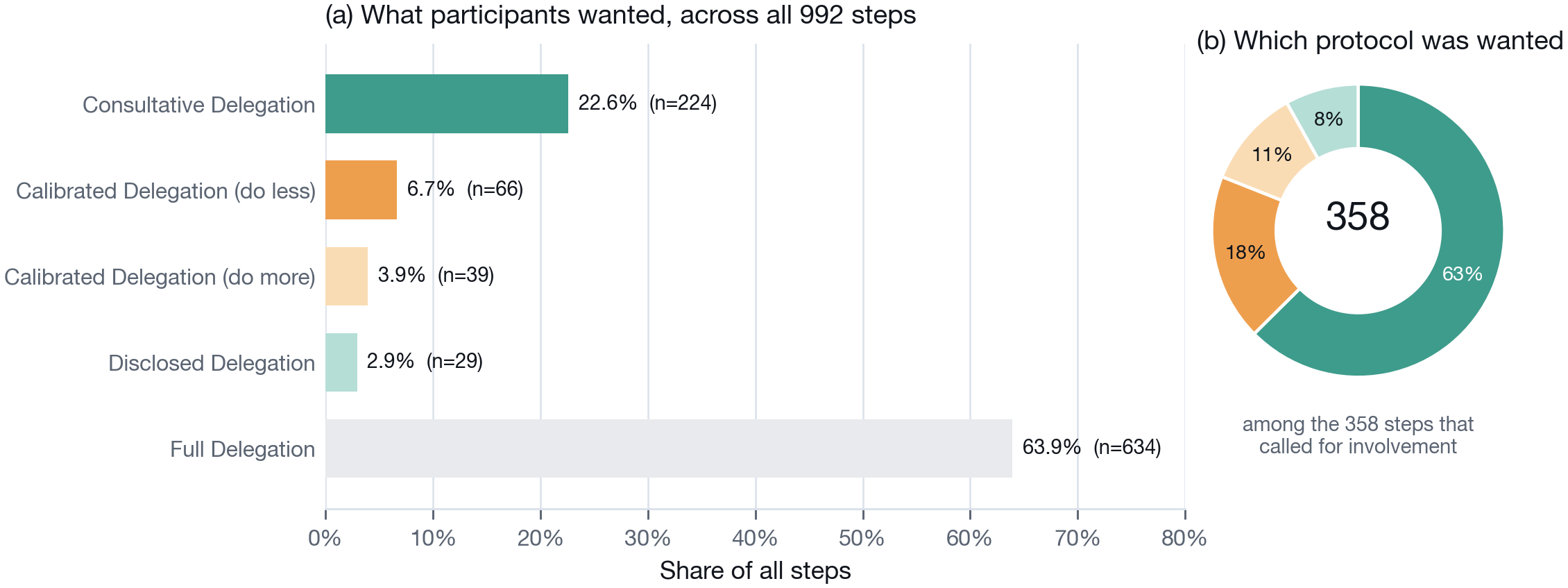}
    \caption{Desired delegation protocols (a) across all steps and (b) among the steps that called for user involvement.}
    \Description{A two-panel chart showing desired delegation protocols. Panel A is a horizontal bar chart across all 992 steps: Full Delegation is most common at 63.9 percent, followed by Consultative Delegation at 22.6 percent, Calibrated Delegation to do less at 6.7 percent, Calibrated Delegation to do more at 3.9 percent, and Disclosed Delegation at 2.9 percent. Panel B is a donut chart for the 358 steps that called for user involvement: 63 percent called for Consultative Delegation, 18 percent for Calibrated Delegation to do less, 11 percent for Calibrated Delegation to do more, and 8 percent for Disclosed Delegation.}
    \label{fig:rq3_protocol}
\end{figure*}
% 1396 words for appendix
\end{document}